\documentclass{article}
\usepackage[margin=1in]{geometry}

\usepackage[numbers]{natbib}

\usepackage{authblk}

\usepackage{amsmath}

\usepackage{moreverb,url}

\usepackage[colorlinks,bookmarksopen,bookmarksnumbered,citecolor=red,urlcolor=red]{hyperref}

\usepackage{multirow}
\usepackage{subcaption}
\usepackage{booktabs}

\usepackage{graphicx}

\newcommand\BibTeX{{\rmfamily B\kern-.05em \textsc{i\kern-.025em b}\kern-.08em
T\kern-.1667em\lower.7ex\hbox{E}\kern-.125emX}}

\DeclareCaptionLabelFormat{addS}{#1 S#2}

\title{Weighting methods for truncation by death in cluster-randomized trials}

\author[1]{Dane Isenberg\thanks{dane.isenberg@pennmedicine.upenn.edu}}
\author[1]{Michael Harhay}
\author[1]{Nandita Mitra\thanks{Co-senior author}}
\author[2,3]{Fan Li$^\dagger$}
\affil[1]{Department of Biostatistics, Epidemiology and Informatics, University of Pennsylvania, Philadelphia, PA, USA}
\affil[2]{Department of Biostatistics, Yale School of Public Health, New Haven, CT, USA}
\affil[3]{Center for Methods in Implementation and Prevention Science, Yale School of Public Health, New Haven, CT, USA}

\begin{document}

\maketitle


\begin{abstract}
Patient-centered outcomes, such as quality of life and length of hospital stay, are the focus in a wide array of clinical studies. However, participants in randomized trials for elderly or critically and severely ill patient populations may have truncated or undefined non-mortality outcomes if they do not survive through the measurement time point. To address truncation by death, the survivor average causal effect (SACE) has been proposed as a causally interpretable subgroup treatment effect defined under the principal stratification framework. However, the majority of methods for estimating SACE have been developed in the context of individually-randomized trials. Only limited discussions have been centered around cluster-randomized trials (CRTs), where methods typically involve strong distributional assumptions for outcome modeling. In this paper, we propose two weighting methods to estimate SACE in CRTs that obviate the need for potentially complicated outcome distribution modeling. We establish the requisite assumptions that address latent clustering effects to enable point identification of SACE, and we provide computationally-efficient asymptotic variance estimators for each weighting estimator. In simulations, we evaluate our weighting estimators, demonstrating their finite-sample operating characteristics and robustness to certain departures from the identification assumptions. We illustrate our methods using data from a CRT to assess the impact of a sedation protocol on mechanical ventilation among children with acute respiratory failure.

\setlength{\parindent}{0pt}\textbf{Keywords}: Estimands, generalized linear mixed models, principal stratification, principal score, survival score, survivor average causal effect, cluster-randomized trials
\end{abstract}

\maketitle

\allowdisplaybreaks

\section{Introduction}

In cluster-randomized trials (CRTs), groups of individuals, such as schools, communities, or health facilities, are randomly assigned to treatments.\citep{murray2004design} CRTs are a favorable alternative to individually-randomized trials when the intervention must be administered to entire groups, there are concerns about contamination (i.e., individuals can gain access to treatment(s) to which they were not assigned), and/or individual random assignment is logistically challenging.\citep{turner2017review,hayes2017cluster} In CRTs, researchers may be interested in studying the causal effect of a binary treatment on a non-mortal outcome like quality of life. However, in some applications, individual participants may die before the time when their non-mortal outcome would be measured, and death acts as an intercurrent event. Mortality rates can be substantial in CRTs among vulnerable populations such as the critically ill or elderly patients. 

Death prior to follow-up is a post-randomization event that precludes the complete measurement of a non-mortal outcome. However, estimation of treatment effects among observed survivors can introduce selection bias when survival status is impacted by treatment because such conditioning disrupts the balance of confounders, measured and unmeasured, afforded by randomization. Bias due to conditioning on post-randomization variables such as death is a well-documented issue in causal inference literature.\citep{rosenbaum1984consequences,wang2017inference,montgomery2018conditioning} To address this problem from a causal perspective, Frangakis and Rubin\citep{frangakis2002principal} proposed the principal stratification framework which partitions populations with respect to cross-classified potential outcomes of a binary post-randomization variable. These latent strata are thus unaffected by treatment assignment and can be considered as pre-treatment covariates that define relevant subpopulations. In the context of truncation by death, a common causal estimand of interest is the mean difference in potential outcomes among individuals who would survive under either treatment received (referred to as the \emph{always-survivors}), named the survivor average causal effect (SACE).\citep{rubin1998more} This is precisely the subpopulation where the pair of non-mortality potential outcomes are well-defined without truncation by death.

For the setting of individually-randomized trials, several methods have been proposed to point identify and estimate SACE. For example, Zhang et al.\citep{zhang2009likelihood} introduced a mixture model approach for empirical identification of SACE, which relies on correctly specifying the joint distribution of principal strata and non-mortal outcomes conditional on covariates. While likelihood inference leads to efficient estimators under correct model assumptions, accurate specification of the conditional outcome models is usually more challenging than the principal strata model. In addition, empirical identification under the mixture model requires sufficient distance between the mixture components and may be numerically less stable otherwise.\citep{mercatanti2015improving} Alternatively, leveraging information from baseline covariates, Hayden et al.\citep{hayden2005estimator} and Ding and Lu\citep{ding2017principal} consider different formulations of ignorability assumptions that allow for non-parametric \textit{point identification} of SACE. These assumptions result in simple weighting methods for these estimands. Employing a similar set of assumptions to Ding and Lu, Zehavi and Nevo\citep{zehavi2023match} analogously develop matching estimators for SACE. Despite the simplicity of the weighting methods, they were designed for independent and identically distributed data and may not be directly applicable to the multilevel data structure present in CRTs. Typically in CRTs, individuals within the same group tend to have positively correlated outcomes, and failing to properly account for this correlation may result in underestimated variances and anti-conservative inference.\citep{turner2017review2} An adjacent problem---one that has a direct impact on causality---is that there may be both observed and unobserved cluster-level confounding if there are cluster characteristics that impact both the survival status and the non-mortal outcome, \citep{li2013propensity} and the implications of the unmeasured cluster-level confounding are unclear in the context of truncation by death in CRTs. 
 
In this paper, we survey the application of weighting estimators for addressing truncation by death in CRTs. We pursue the principal stratification framework and tackle the additional complications associated with the hierarchical data structure imposed by cluster randomization. While likelihood methods have been studied in CRTs to address truncation by death,\citep{tong2023bayesian,he2023bayesian,wang2023mixed} the more intuitive weighting estimators for SACE have not been fully investigated in this context. To dispense with the need to model the outcome distributions, we provide two weighting estimators derived under different sets of identification assumptions inspired by Hayden et al.\citep{hayden2005estimator}, Ding and Lu\citep{ding2017principal}, and Zehavi and Nevo\citep{zehavi2023match}. While the estimators are distinct, they turn out to be functions of the exact same working models to predict survival status, and we provide a conceptual and numerical comparison between these estimators. Furthermore, for each estimator, we consider both conditional and marginal logistic regression paradigms for modeling survival status that are commonly employed in analyses of CRTs.\citep{turner2017review} To accomplish this, we solve for the SACE estimators and construct their sandwich variance expressions using cluster-indexed estimating equations per the M-estimation method\citep{stefanski2002calculus} that are specific to each modeling paradigm. For the conditional model, we fit a generalized mixed model (GLMM) with a random intercept for cluster membership, and for the marginal model, we fit a standard generalized linear model (GLM), which due to said cluster indexing is equivalent to solving a (logit) generalized estimating equation (GEE) with an independence working correlation structure.\citep{liang1986longitudinal} 

We have organized the remaining sections of the article as follows. In Section~\ref{sec:methods}, we define notation, provide two sets of identification assumptions, and present the corresponding weighting estimators for SACE in CRTs, along with their variance estimators. In Section \ref{sec:simulation}, we conduct a simulation study to empirically assess the performance of our estimators under different data generating mechanisms which adhere to each set of assumptions. Moreover, we assess the necessity of explicitly modeling for latent group-level effects by comparing the performance of the conditional to the marginal model for survival status in the SACE estimators. While the proposed marginal model does not account for these effects directly, the cluster-indexed M-estimation results in a sandwich variance estimator that aims to correct for this in inference. In Section~\ref{sec:applications}, we apply our methods to SACE estimation in a CRT that evaluates the impact of an updated sedation protocol on mechanical ventilation duration among children with acute respiratory failure. Section~\ref{sec:discussion} concludes with a discussion. To facilitate implementation of the proposed methods, we provide a companion \texttt{R} package in a GitHub repository\citep{isenberg2024sace} for implementing our weighting estimators in CRTs.

\section{Methods}\label{sec:methods}

\subsection{Notation, Survivor Average Causal Effect Estimand, and Assumptions}

We use the following notation for CRTs. Let $n_c$ denote the number of clusters in a study. The subscript $i$ refers to the $i\text{-th}$ cluster such that $i=1,...,n_c$. Let $n_i < \infty$ be the size of cluster $i$, and $A_i \in \{0,1\}$ be the binary treatment assigned to cluster $i$. The double subscript `$ij$' refers to the $j\text{-th}$ individual, $j=1,...,n_i$, in the $i\text{-th}$ cluster. In the observed data, let $S_{ij} \in \{0,1\}$ be individual survival status before the measurement of the non-mortal outcome, and let $Y_{ij}$ be the non-mortal outcome. Under the potential outcomes framework, we let $S_{ij}(a)$ and $Y_{ij}(a)$ be the potential values of the survival status and non-mortal outcome respectively of individual $j$ in cluster $i$ had cluster $i$ been assigned to $A_i=a$. Potential outcomes are connected to observed outcomes under received treatment assignment through the cluster-level SUTVA (the Stable Unit Treatment Value Assumption), where $S_{ij}=A_{i}S_{ij}(1)+(1-A_{i})S_{ij}(0)$ and if $S_{ij}(a)=1$, then $Y_{ij}=A_{i}Y_{ij}(1)+(1-A_{i})Y_{ij}(0)$. Under truncation by death, if $S_{ij}(a)=0$, then $Y_{ij}(a)=*$ following established notation for SACE.\citep{zhang2003estimation,zhang2009likelihood} In addition, we define $S_i(a)=(S_{i1}(a),...,S_{in_i}(a))^T$, and $Y_i(a)=(Y_{i1}(a),...,Y_{in_i}(a))^T$.

Under the principal stratification framework,\citep{frangakis2002principal} we define the prinicipal strata in terms of the joint variables, $G_{ij}=(S_{ij}(1),S_{ij}(0)) \in \{0,1\} \times \{0,1\}$. These strata partition the population into subgroups defined with post-randomization events (sometimes referred to as intercurrent events) and since they are no longer functions of $A_i=a$, stratum membership can be viewed as a pre-treatment covariate. The strata are \textit{always-survivors} when $G_{ij}=(1,1)$, \textit{protected patients} when $G_{ij}=(1,0)$, \textit{harmed patients} when $G_{ij}=(0,1)$, and \textit{never-survivors} when $G_{ij}=(0,0)$. For studying the treatment effect on the non-mortality outcome, we define the estimand within the sub-population of always-survivors. We restrict the estimand to this stratum because it is the only subgroup for which $Y_{ij}(1)$ and $Y_{ij}(0)$ are both unambiguously defined (or equivalently without impact of the intercurrent event of death). These effects are represented as functions $g(\mu(1),\mu(0))$ where for $a=0,1$,
\begin{equation}\mu(a) =E\{Y_{ij}(a)|G_{ij}=(1,1)\}=\frac{E \{ Y_{ij}(a)S_{ij}(1)S_{ij}(0)\}}{E \{S_{ij}(1)S_{ij}(0)\}}\end{equation}

Possible functions $g(\mu(1),\mu(0))$  may be the mean difference or risk ratio. We primarily focus on the SACE expressed in mean difference (although the methods apply directly to other summary measures) defined as 
\begin{equation}\tau=\mu(1)-\mu(0)=E\{Y_{ij}(1)-Y_{ij}(0)|G_{ij}=(1,1)\}\end{equation}
This estimand represents the expected difference in response had an individual received the active treatment as compared to the status quo treatment given their membership in the sub-population of always-survivors -- that is, a subset of healthier patients for whom the definition of non-mortality potential outcomes is uncomplicated as they are not affected by the intercurrent event of death.
 
Because stratum membership is only partially observed, the estimand $\tau$ cannot be directly identified without additional assumptions. We consider assumptions that require leveraging information from baseline covariates. Let the variables $X_{ij}$ denote vectors of individual-level covariates such that $X_i$ is the covariate matrix for all individuals in cluster $i$, and let the variables $C_i$ denote vectors of cluster-level covariates. It may be that the $C_i$ contain a latent cluster-level variable $b_i$ (assumed to follow a common distribution) that summarizes the impact of unobserved clustering effects and can act as a confounder for the relationship between non-mortal outcome and survival status. In addition to the cluster-level SUTVA, we require that for all $i,j$ and $a=0,1$ that $0<P(S_{ij}(a)=1|X_i,C_i)<1$. In other words, every individual within every cluster has some possibility of either survival or mortality under both treatments. The remaining assumptions are as follows:

{\setlength{\parskip}{10pt} \par \noindent \textbf{Assumption A1}: (\emph{Between-cluster independence}) The potential outcomes and cluster-level baseline variables $\{S_i(1),S_i(0),Y_i(1),Y_i(0),X_i,C_i\}$ are independent across clusters and drawn from a common distribution for a given cluster size. 

{\setlength{\parskip}{10pt} \par \noindent \textbf{Assumption A2}: (\emph{Randomization}) Treatment assignment is independent of all cluster-level variables such that $A_{i} \perp \{S_{i}(1),S_{i}(0),Y_{i}(1),Y_{i}(0),X_i,C_i\}$, and $P(A_i=1)=p \in (0,1)$.}

{\setlength{\parskip}{10pt} \par \noindent \textbf{Assumption A3}: (\emph{Non-informative cluster-size}) Within each cluster $i$, the vector of the individual-level potential outcomes $\{S_{ij}(1),S_{ij}(0),Y_{ij}(1),Y_{ij}(0)\}$ have the same marginal distribution that is independent of the cluster size $n_i$.}

Assumptions A1 and A2 are standard assumptions for CRTs. Assumption A3 prevents the size of each cluster from possibly affecting individuals' survival status and non-mortal outcomes, a phenomenon referred to as informative cluster size.\citep{kahan2023estimands} This assumption thus allows us to define $\mu(a)$ as a ratio of marginal expectations of individual-level potential outcomes without ambiguity. We provide a further discussion of informative cluster size in Section~\ref{sec:discussion}. 
In addition to the aforementioned assumptions, we need one of the following two additional sets of assumptions to identify SACE in CRTs. Set 1 extends the explainable non-random survival assumptions of Hayden et al.\citep{hayden2005estimator}, and Set 2 refers to the Ding and Lu\citep{ding2017principal} and Zehavi and Nevo\citep{zehavi2023match} assumptions based on survival monotonicity and principal ignorability. Both Sets 1 and 2 rely on cross-world assumptions,\citep{richardson2013single} which make claims about survival and/or non-mortal outcome distributions \textit{across} different interventions, possibly conditional on measured and unmeasured variables. The Set 1 assumptions are defined as follows.

{\setlength{\parskip}{10pt} \par \noindent \textbf{Set 1 Assumption A4 (S1A4)}: (\emph{Conditional Survival Independence}) For each individual in a cluster, $S_{ij}(a) \perp S_{ij}(1-a)|X_i,C_i$.}

{\setlength{\parskip}{10pt} \par \noindent \textbf{Set 1 Assumption A5 (S1A5)}: (\emph{Strong Partial Principal Ignorability}) For each individual in a cluster and for $a=0,1$, $Y_{ij}(a) \perp S_{ij}(1-a)|X_i,C_i,\{S_{ij}(a)=1\}$.}

Assumption S1A4 means that an individual's survival status under one treatment is independent of their survival status under the other treatment given measured and potentially unmeasured cluster-level information (the unmeasured information is assumed as the variable $b_i$ that is included in $C_i$). In other words, a patient's survival under one treatment provides no additional knowledge about their survival under the other treatment once a sufficient set of individual-level covariates (such as age, and other baseline clinical characteristics), cluster-level covariates (geographical location such as urban or non-urban), as well as unmeasured cluster-level information (a cluster-level random effect) are identified. The conditional independence statement of S1A5 has a similar interpretation. It means that an individual's non-mortal outcome under one treatment, contingent on surviving to have this recorded, is not informed by their survival status under the other treatment given satisfactory baseline characteristics. To parallel the Set 1 assumptions, the Set 2 assumptions are defined as follows.} 

{\setlength{\parskip}{10pt} \par \noindent \textbf{Set 2 Assumption A4 (S2A4)}: (\emph{Survival Monotonicity}) For each individual in a cluster, $S_{ij}(1) \ge S_{ij}(0)$ such that there is no harmed patients stratum in the study population.}

{\setlength{\parskip}{10pt} \par \noindent \textbf{Set 2 Assumption A5 (S2A5)}: (\emph{Partial Principal Ignorability}) For each individual in a cluster, $Y_{ij}(1) \perp S_{ij}(0)|X_i,C_i,\{S_{ij}(1)=1\}$.}

\begin{figure}
\centering
\captionsetup{labelfont=bf,justification=raggedright,singlelinecheck=false}
\includegraphics[width=.8\textwidth]{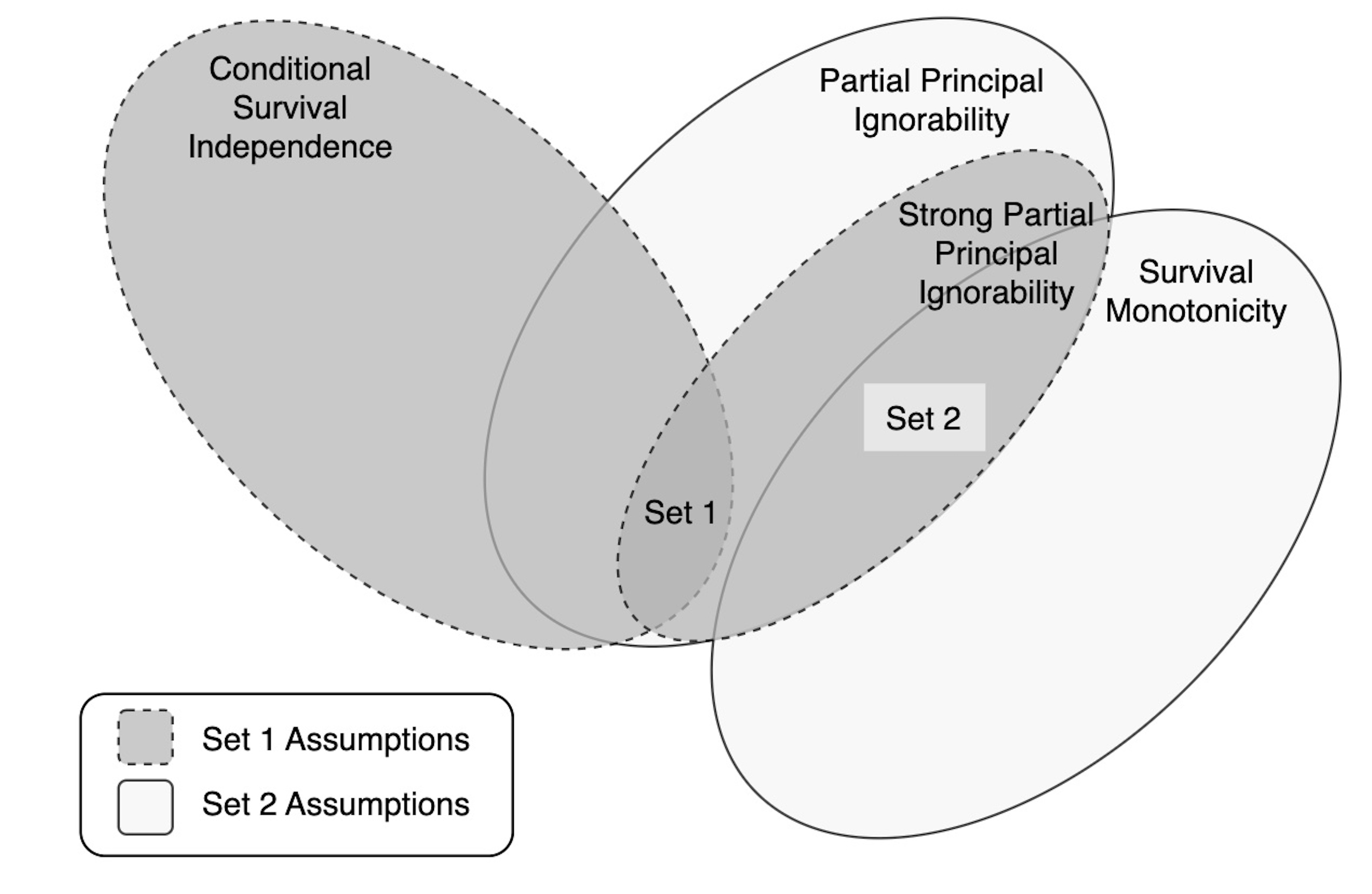}
\caption{Diagram of relationship between Set 1 and Set 2 assumptions for identification of SACE. Each assumption set is satisfied within the intersection of their respective conditions (where they are both preceded by Assumptions 1-3 and SUTVA). S1A4=conditional survival independence, S1A5=strong partial principal ignorability, S2A4=survival monotonicity, S2A5=partial principal ignorability.}
\label{fig:setcon}
\end{figure}

{\setlength{\parskip}{10pt} \par Assumption S2A4 says that survival under the active treatment is no worse than survival under the status quo treatment. S2A5 is a weaker version of S2A5 in that the relationship described need only hold under the active treatment. In their initial conception, Ding and Lu\citep{ding2017principal} provide the stronger assumption of generalized prinicipal ignorability, whose equivalent here would be $Y_{ij}(a) \perp \{S_{ij}(1),S_{ij}(0)\}|X_i,C_i$ for $a=0,1$. The purpose in that more general context is to allow for the identification of estimands defined in additional principal strata, for example, in the presence of treatment noncompliance as an intercurrent event. Under monotonicity, however, it is sufficient to consider S2A5 for identification of the SACE, as pointed out previously.\citep{zehavi2023match}}

As we explain later, each set of assumptions motivate a distinct weighting estimator to identify SACE in CRTs. Given their seeming similarity to one another, it is important to conceptually distinguish between the Set 1 and Set 2 assumptions. We visualize their relationships in Figure~\ref{fig:setcon}, which is meant to demonstrate two salient features. First, S2A4 and S2A5 imply S1A5. For the case of $a=1$, S2A5 directly implies S1A5, and for the case of $a=0$, survival monotonicity, S2A4, means the event $S_{ij}(0)=1$ implies $S_{ij}(1)=1$ (i.e, a constant), which induces the stated conditional independence.\citep{zehavi2023match} Second, the Set 1 and Set 2 assumptions are mutually exclusive due to the relationship between S1A4 and S2A4. If S1A4 and S2A4 are simultaneously true, we would have for each individual and $X_i,C_i$, $P(S_{ij}(1)=0,S_{ij}
(0)=1|X_i,C_i)=P(S_{ij}(1)=0|X_i,C_i)P(S_{ij}(0)=1|X_i,C_i)=0$, which is prohibited by the assumption of non-trivial survival, $0<P(S_{ij}(a)=1|X_i,C_i)<1$, under $a=0,1$. In summary, these two sets of assumptions live disjointly within the condition S1A5. 

\subsection{Two Weighting Estimators}

We define survival probability models as $p^{a}_{ij}(X_i,C_i):=P(S_{ij}=1|A_i=a,X_i,C_i)$, where $P(S_{ij}=1|A_i=a,X_i,C_i)=P(S_{ij}(a)=1|X_i,C_i)$.
At the outset, it may seem as though $\mu(a)$ which conditions on $G_{ij}=(1,1)$ relies on modeling in terms of the principal strata directly. Ding and Lu\citep{ding2017principal} refer to these models of the principal strata probability, $P(G_{ij}= g | X_i,C_i)$, where $g \in \{0,1\} \times \{0,1\}$, as principal score models. However, both these sets of assumptions that enable cross-world comparisons allow for identification of $\mu(a)$ in a way that relies only on modeling survival probabilities. That is, these assumptions demonstrate how we can pass from the joint conditional distribution of survival statuses into modeling them marginally. 

We next present Identity 1 and Identity 2, which represent how SACE is point identified with estimable parameters under the Set 1 and Set 2 assumptions respectively. Importantly, for both, any appearance of the non-mortal outcome will be predicated on survival (due to a product term). Interestingly, these two SACE quantities, while different in form, will depend on the exact same models. 

{\setlength{\parskip}{10pt} \par \noindent \textbf{Identity 1.} The parameters $\mu(a)$ for $a=0,1$ under Set 1 Assumptions are identified with:}
\begin{equation} \label{eq:mua1} \mu(a)=\frac{E \left\{Y_{ij}I(A_i=a)S_{ij}p^{1-a}_{ij}(X_i,C_i)\right\}}{E\left\{I(A_i=a)S_{ij}p^{1-a}_{ij}(X_i,C_i)\right\}}\end{equation}
The weights $I(A_i=a)S_{ij}p_{ij}^{1-a}(X_i,C_i)$ mean that the observed survivors within one treatment group are up-weighted by their probability of survival under the alternative treatment, which in turn signifies that individuals whose probability of being an always-survivor is higher are weighted more heavily. In the context of a cluster-randomized trial, the proof of Identity 1 is provided in Supplemental Materials A.1. 

{\setlength{\parskip}{10pt} \par \noindent \textbf{Identity 2.} The parameters $\mu(a)$ under Set 2 Assumptions are identified with:}  \begin{equation} \label{eq:mu02}\mu(0)=\frac{E \{Y_{ij}(1-A_i)S_{ij} \}}{E \{(1-A_i)S_{ij}\}}\end{equation} and 
\begin{equation}\label{eq:mu12}\mu(1)=\frac{E\left \{{Y_{ij}A_iS_{ij}p^{0}_{ij}(X_i,C_i)}/{p^{1}_{ij}(X_i,C_i)} \right \}}{E\left \{{A_iS_{ij}p^{0}_{ij}(X_i,C_i)}/{p^{1}_{ij}(X_i,C_i)}\right\}}\end{equation}

By cluster-level SUTVA, S2A4, and $A_i \in \{0,1\}$, we have $(1-A_i)S_{ij} =(1-A_i)^2S_{ij}(0)=(1-A_i)S_{ij}(0)S_{ij}(1)$. Therefore, the weights of $\mu(0)$ are straightforward because those who are observed to have survived treatment 0 are always-survivors. The Set 2 assumptions also imply that $p_{ij}^0(X_i,C_i)=P(G_{ij}=(1,1)|X_i,C_i)$. Since $p_{ij}^1(X_i,C_i)=P(G_{ij}=(1,1)|X_i,C_i)+P(G_{ij}=(1,0)|X_i,C_i)$, the terms in $A_iS_{ij}{p^0_{ij}(X_i,C_i)}/{p^1_{ij}(X_i,C_i)}$ of $\mu(1)$ re-weight the observed survivors in the treated group, which are comprised of both always-survivors and protected patients, to reflect the population of always-survivors. In the context of a cluster-randomized trial, the proof of Identity 2 is presented in Supplemental Materials A.2. 

For our working survival modeling, we first  consider that the data, $\{A_i,S_i,C_i,X_i\}$, arise from each cluster independently. Given our beliefs about the mechanism of clustering effects, we will suppose that within each cluster $i$, an individual's survival status is drawn according to $S_{ij}|A_i,X_i,C_i;\gamma \sim \text{Bern}(\text{expit}\{D_{ij}^T\beta+b_i\})$ where $b_i\sim N(0,\sigma^2_b)$ with $\sigma^2_b >0$, and $D_{ij}$ is a finite vector of observed regressors involving $A_i,X_i, C_{i,\text{fe}}$. $C_{i,\text{fe}}$ denotes the component of $C_{i}$ whose values can be observed (as $C_i$ includes the random effect $b_i$ that summarizes the impact of all unobserved cluster-level covariates). In practice, we often make a convenience choice that $D_{ij}=(1,A_i,X_{ij}^T,C_{i,\text{fe}}^T)^T$ such that the survival of each individual only depends on the characteristics of that individual but not those from other individuals in the same cluster (apart from shared cluster-level information). Therefore, for $a=0,1$, each individual's probability of survival is represented as
\begin{equation} \label{eq:glmm} p^{a}_{ij}(X_i,C_i;\gamma)=\frac{\exp\{D_{ij}(a)^T \beta + b_i\}}{1+\exp\{D_{ij}(a)^T \beta + b_i\}}\end{equation} where $D_{ij}(a)$ represents setting $A_i=a$. The effect of treatment and observable baseline covariates on survival is captured in $\beta$, and the unobserved effects of clustering on survival are represented via the random intercept $b_i$ with variance $\sigma^2_b$, so that $\gamma=(\beta^T,\sigma^2_b)^T$ of finite dimension. Although we do not consider this possibility, $D_{ij}$ can also include interaction terms between treatment and covariates as well as some summary measure of cluster-level covariates through $g(X_i)$ (e.g. $\overline{X}_i$) as in a contextual-effects model.\citep{raudenbush1997statistical}

Two common approaches for analyzing survival status data from parallel-arm CRTs -- that we will use for survival status modeling for our SACE estimators -- are to (i) fit a GLMM, specifically a mixed effects logistic regression model with a random intercept for cluster membership and (ii) fit a GLM, specifically a logistic regression model (accounting for clustering only in the variance expression) where both models must include a treatment effect indicator.\citep{murray2004design,turner2017review2} Since the GLMM relies on modeling the distribution of survival status conditional on the random effects (as well as any observed covariates), it is referred to as a conditional model. In contrast, the GLM, for which no random effects are modeled, is referred to as a marginal model under working independence.\citep{heagerty2000marginalized}

\begin{table}
\centering
\captionsetup{labelfont=bf,justification=raggedright,singlelinecheck=false}
\caption{Weighting Estimators for SACE ($\tau$). 
}
\begin{tabular}{l|l}
\toprule
\multicolumn{1}{c|}{Name (Assumptions)} & \multicolumn{1}{c}{Expressions for Estimator} \\
\midrule
$\widehat \tau_{\text{SSW}}$ (Set 1) & $\frac{\displaystyle \sum_{i=1}^{n_c}\sum_{j=1}^{n_i}Y_{ij}A_iS_{ij} p^{0}_{ij}(X_i,C_i;\widehat \gamma)}{\displaystyle \sum_{i=1}^{n_c}\sum_{j=1}^{n_i}A_iS_{ij} p^{0}_{ij}(X_i,C_i;\widehat \gamma)}-\frac{\displaystyle \sum_{i=1}^{n_c}\sum_{j=1}^{n_i}Y_{ij}(1-A_i)S_{ij} p^{1}_{ij}(X_i,C_i;\widehat \gamma)}{\displaystyle \sum_{i=1}^{n_c}\sum_{j=1}^{n_i}(1-A_i)S_{ij} p^{1}_{ij}(X_i,C_i;\widehat \gamma)}$ \\[8ex]
$\widehat \tau_{\text{PSW}}$ (Set 2)& $\frac{\displaystyle  \sum_{i=1}^{n_c}\sum_{j=1}^{n_{i}}   \frac{Y_{ij}A_iS_{ij} p^{0}_{ij}(X_i,C_i;\widehat \gamma)}{p^{1}_{ij}(X_i,C_i;\widehat \gamma)}}{\displaystyle \sum_{i=1}^{n_c}\sum_{j=1}^{n_{i}}\frac{A_iS_{ij} p^{0}_{ij}(X_i,C_i;\widehat \gamma)}{p^{1}_{ij}(X_i,C_i;\widehat \gamma)}}-\frac{\displaystyle \sum_{i=1}^{n_c}\sum_{j=1}^{n_{i}}Y_{ij}(1-A_i)S_{ij}}{\displaystyle \sum_{i=1}^{n_c}\sum_{j=1}^{n_i}(1-A_i)S_{ij}}$ \\ 
\bottomrule
\end{tabular}
\label{tab:saceestim}
\end{table}

We will denote the estimators derived from Set 1 and Set 2, respectively, as the survival score weighting (SSW) estimator, $\widehat \tau_{\text{SSW}}=\widehat \mu_{\text{SSW}}(1)-\widehat \mu_{\text{SSW}}(0)$, and the principal score weighting (PSW) estimator, $\widehat \tau_{\text{PSW}}=\widehat \mu_{\text{PSW}}(1)-\widehat \mu_{\text{PSW}}(0)$. Their forms are summarized in Table~\ref{tab:saceestim}. Although neither estimator fits principal scores directly, Set 2 uses that under monotonocity, the always-survivors principal stratum is comprised of exactly those that survive under treatment 0, whereas Set 1 relies on S1A4 breaking up the principal strata distributions (hence the naming convention). As discussed in Supplemental Materials A.3 and A.4, $\widehat \tau_{\text{SSW}}$ and $\widehat \tau_{\text{PSW}}$ are consistent for SACE if $\widehat \theta_{\text{SSW}}=(\widehat \gamma^T, \widehat \mu_{\text{SSW}}(1),\widehat \mu_{\text{SSW}}(0))^T$ and $\widehat \theta_{\text{PSW}}=(\widehat \gamma^T, \widehat \mu_{\text{PSW}}(1),\widehat \mu_{\text{PSW}}(0))^T$ are each a solution to unbiased estimating equations indexed by cluster, and if under the regularity conditions for which these $\widehat \theta_{(\cdot)} \xrightarrow{p} \theta_0=(\gamma_0^T, \mu(1),\mu(0))^T$ , we have correct specification of the working survival modeling such that $\gamma_0$ reflects the truth.\citep{stefanski2002calculus} If our working model is correctly specified, then to achieve consistency, we would fit the GLMM model (i). 

We consider the fitted GLM model (ii) as the case of a priori setting $\sigma^2_b=0$ (or $b_i=0$ a.s.), so that it is not an estimated parameter. The statements above regarding consistency would remain true under correct specification of the GLM as a predictive model, but we expect as per our working modeling assumptions, that the data contain non-zero $b_i$. Even though the GLM may falsely omit the random intercept term, the cluster-indexed estimating equations result in a sandwich variance expression that accounts for extra variation due to within-group correlation, and it is unclear that this model for prediction always has a negative effect on performance in inference relative to a GLMM.\cite{murray2003methods,campbell2005determinants} This discussion about conditional versus marginal modeling is revisited in greater detail in Section~\ref{sec:varest} and in Supplemental Materials A.6 with relevant citations, and their small-sample performances are directly juxtaposed in the simulation study under data generation that includes a cluster-level random effect.

\subsection{Variance Estimation} \label{sec:varest}

In this section, we approximate the variance of the weighting estimators, $\widehat \tau_{\text{SSW}}$ and $\widehat \tau_{\text{PSW}}$, by establishing their asymptotic distributions through the theory of M-estimation.\citep{stefanski2002calculus} Our motivation for using asymptotic variance expressions stems from the limitations of the cluster bootstrap despite its conceptual simplicity. Specifically, cluster bootstrapping generally requires re-sampling entire clusters with replacement from the original data set\citep{field2007bootstrapping} to preserve the unknown within-cluster correlation structure, so it tends to work better for a large number of clusters.\cite{rabideau2024multiply} However, for even a moderate number of clusters, the computational time for the bootstrap approach with an acceptable number of replicates can be prohibitive since it requires fitting regression models repeatedly (such as GLMM) to each bootstrapped data set. To this end, our proposed variance estimators serve as computationally efficient alternatives to re-sampling-based methods.

Recall we suppose that the available data $\{D_i,S_i,Y_i\}_{i=1}^{n_c}$ are mutually independent by cluster randomization (where the observable elements of $Y_i$ are for observed survivors). Suppose that each $D_{ij}$ and thus $\beta$ have dimension $p$. Our focus is on the space of parameters of the form $\theta=(\gamma^T,\mu(1),\mu(0))^T=(\beta^T,\sigma^2_b,\mu(1),\mu(0))^T$ of dimension $(p+3) \times 1$ since our target estimand $\tau$ is simply the linear combination $k^T \theta$, where $k=(0^T,0,1,-1)^T$. For each $i$, we define a function of observable data, $m(D_i,S_i,Y_i;\theta)=m_{i}(\theta)$, which matches the dimension of $\theta$, such that $E_\theta\{m_{i}(\theta)\}=0$ (for any $\theta$). Then, for the included number of clusters in the study $n_c$, the solution to $\frac{1}{n_c}\sum_{i=1}^{n_c}m_{i}(\theta)=0$ is the M-estimator, $\widehat \theta$. However, since the identification of $\mu(1)$ and $\mu(0)$ are distinct for the Set 1 and Set 2 assumptions, the corresponding unbiased estimating equations must themselves be distinct with functions, $m_{\text{SSW},i}(\theta)$ and $m_{\text{PSW},i}(\theta)$, and solutions, $\widehat \theta_{\text{SSW}}$ and $\widehat \theta_{\text{PSW}}$. 

To obtain the unbiased estimating equations, we stack the score functions used to maximize the (marginalized) log-likelihood induced by the GLMM conditions, $l(S|D,\beta,\sigma^2_b)=\sum_{i=1}^{n_c} l_i(S_i|D_i,\beta,\sigma^2_b)$ (see Supplemental Materials A.4), which will be common to both estimators, and the functions that enable solving for $\widehat \mu_{(\cdot)}(1)$ and $\widehat \mu_{(\cdot)}(0)$ if the GLMM parameters, $\gamma$, were known. The form of estimating functions corresponding to the Set 1 assumptions is \begin{equation} m_{\text{SSW},i}(\theta)=\begin{pmatrix} \displaystyle \partial l_i(S_i|D_i,\beta,\sigma^2_b)/\partial \beta \\ 
\displaystyle \partial l_i(S_i|D_i,\beta,\sigma^2_b)/\partial \sigma^2_b \\  \sum_{j=1}^{n_i}Y_{ij}A_iS_{ij}p^0_{ij}(X_i,C_i;\gamma)-\mu(1)\sum_{j=1}^{n_i}A_iS_{ij}p^0_{ij}(X_i,C_i;\gamma)\\
 \sum_{j=1}^{n_i}Y_{ij}(1-A_i)S_{ij}p^1_{ij}(X_i,C_i;\gamma)-\mu(0)\sum_{j=1}^{n_i}(1-A_i)S_{ij}p^1_{ij}(X_i,C_i;\gamma)\end{pmatrix}\end{equation}
and for the Set 2 assumptions
 \begin{equation} m_{{\text{PSW}},i}(\theta)=\begin{pmatrix}  {\partial l_i(S_i|D_i,\beta,\sigma^2_b)}/{\partial \beta} \\ 
\displaystyle \partial l_i(S_i|D_i,\beta,\sigma^2_b)/\partial \sigma^2_b \\ 
\sum_{j=1}^{n_i} \frac{Y_{ij}A_iS_{ij} p^{0}_{ij}(X_i,C_i;\gamma)}{ p^{1}_{ij}(X_i,C_i;\gamma)}-\mu(1)\sum_{j=1}^{n_i}   \frac{A_iS_{ij}p^{0}_{ij}(X_i,C_i)}{p^{1}_{ij}(X_i,C_i;\gamma)}\\
 \sum_{j=1}^{n_i}Y_{ij}(1-A_i)S_{ij}-\mu(0)\sum_{j=1}^{n_i}(1-A_i)S_{ij}\end{pmatrix} \end{equation}

Using first-order Taylor approximations coupled with certain regularity conditions for these independent data, applications of Slutsky's and continuous mapping theorems give us that $\sqrt{n_c}(\widehat \tau_{\text{SSW}}-\tau_0)$ and $\sqrt{n_c}(\widehat \tau_{\text{PSW}}-\tau_0)$ converge in distribution to $N(0,k^TV_{\theta_0}k)$.\citep{stefanski2002calculus} Under correct specification, $\tau_0$ and $\theta_0$ are the true parameters, and $V_{\theta_0}$ is the true asymptotic variance as written out in Supplemental Materials A.3  (Equation 36). Note, we only subscript estimators with SSW and PSW and not parameters because parameters cannot be simultaneously identified under Set 1 and Set 2. The form of the estimator of the asymptotic variance, often termed a cluster-robust sandwich variance estimator, is:
\begin{equation}\label{eq:sandvar}
\widehat V_{\widehat \theta_{(\cdot)}}=\left[\frac{1}{n_c}\sum_{i=1}^{n_c} \left \{\frac{\partial m_{(\cdot),i}(\widehat \theta_{(\cdot)})}{\partial \theta^T}\right \} \right]^{-1} \frac{1}{n_c}\sum_{i=1}^{n_c}m_{(\cdot),i}(\widehat \theta)m_{(\cdot),i}(\widehat \theta)^T  \left[\frac{1}{n_c} \sum_{i=1}^{n_c} \left \{\frac{\partial m_{(\cdot),i}(\widehat \theta_{(\cdot)})}{\partial \theta^T} \right \}\right]^{-T}
\end{equation}
The regularity conditions under which $\widehat V_{
\widehat \theta_{(\cdot)}}$ is consistent for $V_{\theta_0}$ are posited by Iverson and Randles\citep{iverson1989effects}. The full expressions for the estimating functions (including the derivative of the log-likelihood) and a more thorough sketch of the asymptotic arguments are in Supplemental Materials A.3 and A.4. 

Now, we must determine the components of the sandwich variance for each estimator. We denote the middle matrices with $M_{\text{SSW}}(\theta)=\sum_{i=1}^{n_c}m_{\text{SSW},i}(\theta)m_{\text{SSW},i}(\theta)^T$ and $M_{\text{PSW}}(\theta)=\sum_{i=1}^{n_c}m_{\text{PSW},i}(\theta)m_{\text{PSW},i}(\theta)^T$. The estimating functions that are required to find the middle matrices are already written above. We must also find the outer matrices, which we will denote by $B_{\text{SSW}}(\theta)=\sum_{i=i}^{n_c}\partial m_{\text{SSW},i}(\theta)/\partial \theta^T$ and $B_{\text{PSW}}(\theta)=\sum_{i=i}^{n_c}\partial m_{\text{PSW},i}(\theta)/\partial \theta^T$. The general form of the outer matrices, $B_{(\cdot)}(\theta)$, is:
\begin{equation}
\begin{pmatrix}
\displaystyle \frac{\partial^2 l(S|D,\beta,\sigma^2_b)}{\partial \beta^T \partial \beta} & \displaystyle \frac{\partial^2 l(S|D,\beta,\sigma^2_b)}{\partial \sigma^2_b \partial \beta } & \displaystyle \frac{\partial^2 l(S|D,\beta,\sigma^2_b)}{\partial \mu(1) \partial \beta} & \displaystyle \frac{\partial^2 l(S|D,\beta,\sigma^2_b)}{\partial \mu(0) \partial \beta}\\
\displaystyle \frac{\partial^2 l(S|D,\beta,\sigma^2_b)}{\partial \beta^T \partial \sigma^2_b} &\displaystyle  \frac{\partial^2 l(S|D,\beta,\sigma^2_b)}{\partial (\sigma^2_b)^2} & \displaystyle  \frac{\partial^2 l(S|D,\beta,\sigma^2_b)}{\partial \mu(1) \partial \sigma^2_b} & \displaystyle  \frac{\partial^2 l(S|D,\beta,\sigma^2_b)}{\partial \mu(0) \partial \sigma^2_b}\\
\displaystyle \sum_{i=1}^{n_c} \frac{\partial m_{(\cdot),i,p+2}(\theta)}{\partial \beta^T} & \displaystyle \sum_{i=1}^{n_c} \frac{\partial m_{(\cdot),i,p+2}(\theta)}{\partial \sigma^2_b} & \displaystyle \sum_{i=1}^{n_c} \frac{\partial m_{(\cdot),i,p+2}(\theta)}{\partial \mu(1)} & \displaystyle \sum_{i=1}^{n_c}\frac{\partial m_{(\cdot),i,p+2}(\theta)}{\partial \mu(0)}\\
\displaystyle \sum_{i=1}^{n_c}\frac{\partial m_{(\cdot),i,p+3}(\theta)}{\partial \beta^T} & \displaystyle \sum_{i=1}^{n_c}\frac{\partial m_{(\cdot),i,p+3}(\theta)}{\partial \sigma^2_b} & \displaystyle \sum_{i=1}^{n_c}\frac{\partial m_{(\cdot),i,{p+3}}(\theta)}{\partial \mu(1) } & \displaystyle \sum_{i=1}^{n_c}\frac{\partial m_{(\cdot),i,{p+3}}(\theta)}{\partial \mu(0)}
\end{pmatrix}\end{equation}
where the third index of $m_{(\cdot),i,k}(\theta)$ represents the $k$-th component of the $i$-th vector function. Due to the nature of the stacked estimating equations, for which the score equations are solved independently, we have matrix structures $B_{\text{SSW}}(\theta)$ and $B_{\text{PSW}}(\theta)$ that are relatively sparse. However, they have different forms since $\mu(0)$ under the Set 2 assumptions is not identified with a parameter that involves a model for survival events. For the SSW estimator, we have outer matrix:
\begin{equation}B_{\text{SSW}}(\theta)=\begin{pmatrix} 
B^{p \times p}_{11} & B^{p \times 1}_{12} & 0^{p \times 1} & 0^{p \times 1}\\
B^{1 \times p}_{21} &  B^{1 \times 1}_{2 2}& 0^{1 \times 1} & 0^{1 \times 1}\\
B^{1 \times p}_{\text{SSW},31} & 0^{1 \times 1} & B^{1 \times 1}_{\text{SSW},33}& 0^{1 \times 1}\\
B^{1 \times p}_{\text{SSW},41} & 0^{1 \times 1} & 0^{1 \times 1} & B^{1 \times 1}_{\text{SSW},44}
\end{pmatrix}\end{equation}
for the PSW estimator,
\begin{equation}
B_{\text{PSW}}(\theta)=\begin{pmatrix}
B^{p \times p}_{11} & B^{p \times 1}_{12} & 0^{p \times 1} & 0^{p \times 1}\\
 B^{1 \times p}_{21} &  B^{1 \times 1}_{2 2}& 0^{1 \times 1} & 0^{1 \times 1}\\
B^{1 \times p}_{\text{PSW},31} & 0^{1 \times 1} & B^{1 \times 1}_{\text{PSW},33}& 0^{1 \times 1}\\
0^{1 \times p} & 0^{1 \times 1} & 0^{1 \times 1} & B^{1 \times 1}_{\text{PSW},44}
\end{pmatrix}\end{equation}
where the superscripts denote the dimension of the submatrices ($\theta$ on the RHS is suppressed for notation simplicity). Using the estimator of the asymptotic variance as in Equation~\eqref{eq:sandvar}, we can approximate the variances for the SACE estimators, $\widehat \tau_{\text{SSW}}$ and $\widehat \tau_{\text{PSW}}$, with $k^T B_{\text{SSW}}(\widehat \theta_{\text{SSW}})^{-1} M_{\text{SSW}}(\widehat \theta_{\text{SSW}}) B_{\text{SSW}}(\widehat \theta_{\text{SSW}})^{-T} k$ and $k^T B_{\text{PSW}}(\widehat \theta_{\text{PSW}})^{-1} M_{\text{PSW}}(\widehat \theta_{\text{PSW}}) B_{\text{PSW}}(\widehat \theta_{\text{PSW}})^{-T} k$, respectively. These expressions assume that $ B_{\text{SSW}}(\widehat \theta_{\text{SSW}})$ and  $B_{\text{PSW}}(\widehat \theta_{\text{PSW}})$ are non-singular, which will generally be the case for sufficient size $n_c$. Each non-zero entry of $B_{\text{SSW}}(\theta)$ and $B_{\text{PSW}}(\theta)$ is fully written out in Supplemental Materials A.4. Estimates of the GLMM parameters can be obtained from standard statistical software. However, both the outer and middle matrices involve integrals, as a byproduct of marginalizing over the random intercept in the likelihood, that are not solvable analytically. Computing these integrals numerically is not a trivial exercise as shown in prior work.\citep{wu2019model} Since the entries in the matrices are complex (particularly in the outer matrix), keeping track of the integral expressions and ensuring that they are numerically integrable is a cumbersome task. After accounting for these integrals, we employed a numerical integration technique that is well-suited to integrals representing Gaussian expectations. In the Supplemental Materials A.5, we present all integrals involved in variance estimation as well as outline our choice and usage of Adaptive Gauss-Hermite Quadrature (AGQ) \citep{liu1994note} for the required numerical integration. Despite the necessity of said numerical methods, for the cases in our simulation study where the non-parametric cluster bootstrap is viable, the variance estimates using the asymptotic distributions are not meaningfully different than those using the bootstrap but the computation times are reduced substantially (for reference, see Supplemental Materials B Table S4). 

The process for determining asymptotic distributions under the GLM model for survival status is a special case of the above where the terms involving $\sigma^2_b$ in the estimating functions are omitted since its value is preset to 0; accordingly, this case is outlined in the Supplemental Materials A.6 and requires no numerical integration (an even greater boon to computation time relative to bootstrapping). Despite this simplification, we stress marginal and conditional logistic regression models are separate conceptual frameworks that also perform differently with clustered data for a binary response (due to a lack of collapsibility) \citep{neuhaus1991comparison} as illustrated in the simulation study in Section~\ref{sec:simulation}. As such, for clarity of notation, we will heretofore distinguish the GLM model from the GLMM by using parameter estimators denoted with a tilde, which so far include: $\widetilde \beta$, $\widetilde \theta_{\text{SSW}}$, $\widetilde \mu_{\text{SSW}}(a)$, $\widetilde \tau_{\text{SSW}}$, $\widetilde \theta_{\text{PSW}}$, $\widetilde \mu_{\text{PSW}}(a)$, and $\widetilde \tau_{\text{PSW}}$ for $a=0,1$. We note that Table \ref{tab:saceestim} omits this notation for continuity and succinctness. It is also important to emphasize that the parameter estimates for SACE are conceived as solutions to estimating equations that sum with respect to the \textit{cluster index}. Therefore, even if the random intercept term is dropped \`a la the marginal GLM model, we obtain a sandwich variance expression that accounts for clustering. More specifically, the sub-matrix of $ B_{(\cdot)}(\widetilde \theta_{(\cdot)})^{-1} M_{(\cdot)}(\widetilde \theta_{(\cdot)}) B_{(\cdot)}(\widetilde \theta_{(\cdot)})^{-T}$ corresponding to the estimated variance matrix for the fixed effects coefficients is precisely the sandwich variance estimate that we would achieve by solving a GEE  \citep{liang1986longitudinal} with mean and variance models as in logistic regression with a working correlation matrix for independence. Solving GEEs are often thought of as the marginal model counterpart to fitting conditional GLMM models,\citep{murray2004design,turner2017review2,lee2004conditional,muff2016marginal} so we find it instructive to underscore the equivalence of our method to solving a GEE. In truth, there is really a coincidence at play here, which is that solving the score equations for the parametric GLM is equivalent to the semiparametric method of solving the logit GEE with an independence correlation structure -- analogous to the equality of the MLE to the least-squares solution for linear regression. As a result, this sandwich variance is often (aptly) referred to as the cluster-robust variance estimate under the GLM, and it is why it can be perceived as a post-hoc correction for variance underestimation under an iid assumption.\citep{cameron2015practitioner} In the Supplemental Materials A.6, we illustrate the equivalence of our approach to solving a GEE in the course of providing the stacked estimating equations that enable estimation of $\widetilde \tau_{\text{SSW}}$ and $\widetilde \tau_{\text{PSW}}$.

There are well-documented issues with small-sample inference associated with sandwich variance estimators, either from marginal or conditional modeling, for the analysis of CRTs. These estimators tend to underestimate the true variance in small samples and thus lead to anticonservative inference.\citep{li2015small,maas2004robustness,huang2022accounting,rodriguez1995assessment,kahan2016increased}
As such, we suggest employing a bias-corrected version of the proposed sandwich variance estimators (whether or not the random effect is removed). For our simulation study and data application, we use a simple degrees of freedom correction,\citep{mackinnon1985some} multiplying the variance terms by $n_c/(n_c-\#\text{params})$, where for the GLMM we have one additional parameter for the random effect variance. 

Lastly, it may be that for the fitted GLMM model, $\sigma^2_b$ is estimated to be 0. This estimate arises due to the fact that convergence to a maximum is restricted by the boundary condition on the parameter space, where the variance of the random intercept must be at least 0; for further treatment of this phenomenon including asymptotics, see Self and Liang\citep{self1987asymptotic}. Therefore, constrained optimization algorithms will permit estimates of 0 for the variance, which is the default for the prevailing GLMM fitting package \texttt{lme4} in \texttt{R}.\citep{bates2009package} When this phenomenon occurs, it produces (virtually) the same coefficient estimates as the GLM, and thus can be constructed as the solution to the same estimating equations (i.e., those which drop the random intercept) \citep{barr2013random}; nonetheless, we treat it as a separate case from the marginal model approach because we penalize for the extra variance parameter. There are options to explicitly avoid a zero variance estimate such as changing the optimizer or putting a prior on the variance parameter that pulls it away from 0,\citep{chung2013nondegenerate} but we adopt the most standard application of the existing \texttt{R} software.

\section{Simulation Study}\label{sec:simulation}

In our simulation study, we assess the performance of the SACE estimators under different data generating mechanisms. We follow the ADEMP (Aims, Data-generating mechanisms, Methods, Estimands, Performance measures) schema defined by Morris et al.\citep{morris2019using} to describe our simulation study. There are two primary goals for these simulations. The first goal is to evaluate the two weighting SACE estimators when the data generation adheres to either the Set 1 assumptions or the Set 2 assumptions in terms of their empirical bias and coverage (see Section~\ref{sec:simresult} for more details). We recall that these are technically mutually exclusive data generating mechanisms because S1A4, conditional survival independence, and S2A4, the deterministic requirement of survival monotonicity, cannot exist simultaneously. Because of S1A4, it is natural to define a data generating mechanism that adheres to the Set 1 assumptions but operates on the survival terms separately. However, without generating the principal strata membership directly, it is not straightforward how one achieves monotonicity under Set 2 while also permitting this separation of survival event models. We follow the approach of Jiang et al.\citep{jiang2022multiply} by imposing \textit{stochastic} monotonicity \citep{small2014instrumental}, where we make the probability of survival under treatment $a=1$ greater than under treatment $a=0$ conditional on covariates. For our simulation, we increase the magnitude of the treatment effect on survival substantially so as to drive the empirical incidence of harmed patients to near 0. In some sense, this accords the most with intuition given that monotonicity is achieved under the presumption that the active treatment is effective in promoting survival. Another option for simulating survival monotonicity that affords this separate survival modeling is by treating stratum membership as an ordinal variable using a latent continuous response framework with cut-points that represent the treatment effect on survival. While deterministic monotonicity is achieved this way, the choice of representing stratum membership ordinally is not intrinsic. Additionally, the results are consistent relative to the other data generating mechanisms, so we leave this exploration in Supplemental Materials A.7. The second aim is to evaluate how well our SACE estimators handle latent clustering effects, represented by unobserved cluster-level random effects in the data generating process for the mortality and non-mortality outcomes. To this end, we compare the performance of the estimators with GLMM survival status modeling, $\widehat \tau_{\text{SSW}}$ and $\widehat \tau_{\text{PSW}}$, to those with GLM survival status modeling (with cluster-robust variance), $\widetilde \tau_{\text{SSW}}$ and $\widetilde \tau_{\text{PSW}}$. Variance expressions and confidence z-intervals are computed based on the asymptotic distributions discussed in Section~\ref{sec:methods} and in Supplemental Materials A.

\subsection{Data Generating Process}

We consider scenarios for which we choose the number of clusters $n_c\in\{30,60,90\}$. The number of individuals in each cluster is drawn from a discrete uniform distribution, $n_i \sim \text{Unif}\{25,50\}$. We generate two individual-level covariates for each cluster, $X_{i,1} \sim \text{MVN}_{n_i}(\mathbf{2},0.5I_{n_i})$ and $X_{i,2} \sim \text{MVN}_{n_i}(\mathbf{.5},0.25I_{n_i})$ and one cluster-level covariate, $C_{i,1} \sim \text{Bern}(0.3)$. We incorporate latent cluster-level effects via the variables $b_i$ and $b_i^*$ for survival status and the non-mortal outcome respectively, where we suppose $b^{*}_i \sim N \left(0,1/9\right)$ and $b_{i} \sim \xi b^{*}_i$; this setup also induces unobserved confounding of the relationship between the non-mortal and survival events. These variables differ from the observed covariates because we only propose their distributional form and relationship to other variables, but their actual values are unknown to us. For $a=0,1$, we generate survival status and non-mortal outcomes according to: 
\begin{equation}S_{ij}(a)|X_{ij},C_{i,1},b_{i} \sim \text{Bern}(\text{expit}\{0.75+\delta a+0.1X_{ij,1}-0.05X_{ij,2}+0.1C_{i,1}+b_{i}\})\end{equation} \begin{equation}Y_{ij}(a)|S_{ij}(a)=1,X_{ij},C_{i,1},b^*_{i} \sim \text{N}\left((a+1)(1+0.25X_{ij,1}+0.125X_{ij,2})+b^*_{i},1\right)\end{equation}
For treatment effect, we consider $\delta\in\{0,\log(1.25),\log(5)\}$. The values of $0$ and $\log(1.25)$ represent Set 1 data generation and $\log(5)$ represents Set 2 data generation because the treatment effect is strong enough to induce near empirical monotonocity.

We quantify the strength of the latent clustering effects for each data generating mechanism via the intracluster correlation coefficient (ICC). For the continuous non-mortal outcomes, the ICC or equivalently the correlation among individuals within the same cluster is given by $\rho = \sigma^2_{b^*}/(\sigma^2_{b^*}+\sigma^2_{\epsilon})$, which is set equal to $0.1$. We define the ICC for survival as $\lambda = \sigma^2_b/(\sigma^2_b+\pi^2/3)$. This standard representation for ICC for mixed effects logistic regression is motivated via a latent variable model, which creates a dichotomous outcome by thresholding an underlying continuous variable with a standard logistic distribution.\citep{goldstein2002partitioning,eldridge2009intra} We briefly show how it is derived as a note in Supplemental Materials A.7. This definition of the ICC for GLMMs then allows for an analogous interpretation to the ICC for linear mixed models, except that it is defined on the latent or log-odds scale. Since $b_i \sim N(0,\xi^2 \sigma^2_{b^*})$, we choose values of $\xi$ such that $\lambda=\xi^2 \sigma^2_{b^*}/(\xi^2 \sigma^2_{b^*}+\pi^2/3) \in \{0.3, 0.1\}$. 

For finding the true parameters, we generate the data using a large number of clusters, $n_c=1000$, as if to represent the population of clusters. 
We define the true SACE estimand within each simulation as 
\begin{equation}\frac{\sum_{i=1}^{1000}\sum_{j=1}^{n_i}Y_{ij}(1)S_{ij}(0)S_{ij}(1)}{\sum_{i=1}^{1000}\sum_{j=1}^{n_i}S_{ij}(0)S_{ij}(1)}-\frac{\sum_{i=1}^{1000}\sum_{j=1}^{n_i}Y_{ij}(0)S_{ij}(0)S_{ij}(1)}{\sum_{i=1}^{1000}\sum_{j=1}^{n_i}S_{ij}(0)S_{ij}(1)}\end{equation}
For the results presented in the next section, the percentage of always-survivors under $n_c=1000$ ranges from $51\%$ to $65\%$, where larger values of $\delta$ and smaller $\lambda$ induce higher rates of this subgroup.

For each data generating mechanism, the observed data are obtained after random allocation of clusters to treatment where $A_i \sim \text{Bern}(.5)$. For example, if cluster $i$ is assigned to treatment 1, the observed data are $\{Y_{i}(1),S_{i}(1),X_{i},C_{i,1}\}$ with cluster size $n_i$. In the next section, we report the performance of SACE estimates on these observed data under the different data generation assumptions and modeling techniques as described above. The results for the additional values of $\delta=-\log(1.25)$ and $\lambda\in\{0.15,0.05\}$ as well as the scenarios for deterministic monotonicity are in Supplemental Materials B. 

\subsection{Simulation Results} \label{sec:simresult}

We include four measures to evaluate the performance of the estimators under the various simulation scenarios: empirical bias, empirical variance, average of the estimated variance, and coverage of the $95\%$ confidence interval. The empirical or Monte Carlo variance is included in parentheses as a benchmark for the average of the estimated variance. Results for 1000 simulations are presented in Table~\ref{tab:simresults}, where the bias and variance metrics are scaled by 100. 

\begin{table}
    \centering
\captionsetup{labelfont=bf,justification=raggedright,singlelinecheck=false}
    \renewcommand{\arraystretch}{0.8} 
    \caption{Comparison of average point estimates and variance estimates as well as performance measures of bias ($\times 10^{-2}$) and coverage ($\%$) for nominally $95\%$ z-intervals of the SSW and PSW estimators. Changes to cluster size $n_c$, treatment effects $\delta$, survival ICC $\lambda$, and modeling choices, GLMM (RE=T) vs. GLM (RE=F), are considered. Variances are computed using a degrees of freedom correction and are compared to the empirical variance (EV) in parentheses ($\times 10^{-2}$). 1000 simulations are run according to the proposed data generating mechanisms. $^{\dagger}$ indicates the setting of empirical monotonicity, and $^{*}$ indicates value is scaled up by $100$.}
    \begin{tabular}{ccc|cc|cc|cc|cc}
    \toprule
        \multicolumn{11}{c}{$\lambda=0.1$}\\
    \midrule
        \multicolumn{3}{c|}{Scenarios} & \multicolumn{2}{c|}{Estimates} & \multicolumn{2}{c|}{Model Variance$^{*}$ (EV$^{*}$)} & \multicolumn{2}{c|}{Bias$^{*}$} & \multicolumn{2}{c}{$\%$ Coverage}\\
        $\delta$ & $n_c$ & RE & SSW & PSW & SSW & PSW & SSW & PSW & SSW & PSW\\
    \midrule
        \multirow{6}{*}{0.0} & \multirow{2}{*}{30} & T & 1.567 & 1.564 & 2.7 (2.4) & 2.6 (2.2) & -0.1 & -0.4 & 95.1 & 95.8 \\ 
        & & F & 1.567 & 1.564 & 2.4 (2.1) & 2.4 (2.1) & -0.1 & -0.4 & 95.4 & 95.3 \\ 
        & \multirow{2}{*}{60} & T & 1.564 & 1.562 & 1.2 (1.2) & 1.1 (1.1) & -0.4 & -0.6 & 93.1 & 93.9 \\ 
        & & F & 1.566 & 1.563 & 1.1 (1.1) & 1.1 (1.1) & -0.2 & -0.5 & 94.3 & 94.6 \\ 
        & \multirow{2}{*}{90} & T & 1.566 & 1.563 & 0.7 (0.7) & 0.7 (0.7) & -0.2 & -0.5 & 94.9 & 95.2 \\ 
        & & F & 1.566 & 1.563 & 0.7 (0.6) & 0.7 (0.6) & -0.2 & -0.5 & 95.3 & 95.3 \\ 
        \midrule
        \multirow{6}{*}{0.2} & \multirow{2}{*}{30} & T & 1.564 & 1.562 & 2.6 (2.3) & 2.6 (2.2) & -0.4 & -0.6 & 95.4 & 96.1 \\ 
        & & F & 1.560 & 1.558 & 2.4 (2.1) & 2.4 (2.1) & -0.7 & -1.0 & 95.3 & 95.4 \\ 
        & \multirow{2}{*}{60} & T & 1.561 & 1.560 & 1.1 (1.2) & 1.1 (1.1) & -0.7 & -0.7 & 93.7 & 94.1 \\ 
        & & F & 1.558 & 1.555 & 1.1 (1.1) & 1.1 (1.1) & -0.9 & -1.2 & 94.5 & 94.6 \\ 
        & \multirow{2}{*}{90} & T & 1.562 & 1.560 & 0.7 (0.7) & 0.7 (0.7) & -0.5 & -0.7 & 95.2 & 94.6 \\ 
        & & F & 1.558 & 1.555 & 0.7 (0.6) & 0.7 (0.6) & -1.0 & -1.2 & 94.7 & 94.7 \\ 
        \midrule
        \multirow{6}{*}{1.6$^{\dagger}$} & \multirow{2}{*}{30} & T & 1.541 & 1.543 & 2.7 (2.1) & 2.6 (2.1) & -2.5 & -2.4 & 95.6 & 96.2 \\ 
        & & F & 1.528 & 1.528 & 2.4 (2.1) & 2.4 (2.0) & -3.8 & -3.8 & 94.7 & 94.7 \\ 
        & \multirow{2}{*}{60} & T & 1.541 & 1.543 & 1.2 (1.1) & 1.1 (1.1) & -2.5 & -2.3 & 94.1 & 94.6 \\ 
        & & F & 1.526 & 1.525 & 1.1 (1.0) & 1.1 (1.0) & -4.0 & -4.1 & 92.5 & 92.6 \\ 
        & \multirow{2}{*}{90} & T & 1.542 & 1.543 & 0.7 (0.7) & 0.7 (0.7) & -2.4 & -2.3 & 94.8 & 95.1 \\ 
        & & F & 1.526 & 1.525 & 0.7 (0.6) & 0.7 (0.6) & -4.0 & -4.1 & 92.9 & 92.8 \\ 
        \toprule
        \multicolumn{11}{c}{$\lambda=0.3$}\\
        \midrule
        \multirow{6}{*}{0.0} & \multirow{2}{*}{30} & T & 1.568 & 1.567 & 2.5 (2.6) & 2.5 (2.4) & 0.1 & 0.0 & 93.1 & 93.4 \\ 
        & & F & 1.569 & 1.566 & 2.3 (2.0) & 2.3 (2.0) & 0.2 & -0.1 & 94.5 & 94.1 \\ 
        & \multirow{2}{*}{60} & T & 1.562 & 1.560 & 1.1 (1.4) & 1.1 (1.2) & -0.5 & -0.7 & 91.1 & 93.2 \\ 
        & & F & 1.562 & 1.560 & 1.0 (1.0) & 1.0 (1.0) & -0.5 & -0.7 & 94.4 & 94.6 \\ 
        & \multirow{2}{*}{90} & T & 1.566 & 1.563 & 0.7 (0.8) & 0.7 (0.7) & -0.1 & -0.4 & 92.2 & 93.2 \\ 
        & & F & 1.567 & 1.564 & 0.6 (0.6) & 0.6 (0.6) & -0.1 & -0.4 & 94.5 & 94.5 \\ 
        \midrule
        \multirow{6}{*}{0.2} & \multirow{2}{*}{30} & T & 1.566 & 1.567 & 2.5 (2.6) & 2.5 (2.4) & 0.0 & 0.0 & 93.0 & 93.8 \\ 
        & & F & 1.559 & 1.557 & 2.3 (2.0) & 2.3 (2.0) & -0.7 & -1.0 & 94.3 & 94.0 \\ 
        & \multirow{2}{*}{60} & T & 1.560 & 1.559 & 1.1 (1.3) & 1.0 (1.2) & -0.6 & -0.8 & 91.0 & 92.2 \\ 
        & & F & 1.552 & 1.550 & 1.0 (1.0) & 1.0 (1.0) & -1.5 & -1.7 & 94.1 & 94.0 \\ 
        & \multirow{2}{*}{90} & T & 1.564 & 1.562 & 0.7 (0.8) & 0.7 (0.7) & -0.3 & -0.5 & 92.8 & 93.2 \\ 
        & & F & 1.555 & 1.553 & 0.6 (0.6) & 0.6 (0.6) & -1.2 & -1.4 & 94.5 & 94.4 \\ 
        \midrule
        \multirow{6}{*}{1.6$^{\dagger}$} & \multirow{2}{*}{30} & T & 1.549 & 1.549 & 2.5 (2.3) & 2.5 (2.2) & -1.7 & -1.7 & 94.2 & 94.5 \\ 
        & & F & 1.508 & 1.507 & 2.2 (1.9) & 2.2 (1.9) & -5.9 & -5.9 & 93.2 & 93.1 \\ 
        & \multirow{2}{*}{60} & T & 1.546 & 1.547 & 1.1 (1.2) & 1.1 (1.1) & -2.0 & -1.8 & 92.8 & 93.1 \\ 
        & & F & 1.503 & 1.502 & 1.0 (1.0) & 1.0 (1.0) & -6.3 & -6.3 & 89.6 & 89.2 \\ 
        & \multirow{2}{*}{90} & T & 1.549 & 1.549 & 0.7 (0.7) & 0.7 (0.7) & -1.7 & -1.6 & 93.5 & 93.6 \\ 
        & & F & 1.504 & 1.503 & 0.6 (0.6) & 0.6 (0.6) & -6.2 & -6.3 & 87.7 & 87.5 \\ 
    \bottomrule
     \end{tabular}
    \label{tab:simresults}
\end{table}

Overall, the simulation study reveals that both of the proposed SACE estimators, where the mortality outcome is modeled using mixed effects logistic regression, exhibit low bias across all data generating scenarios. The $95\%$ z-intervals constructed using the asymptotic variance expressions with a degrees of freedom adjustment in general get close to nominal rates. In summary, regardless of the assumptions underlying the data generating mechanism, $\widehat \tau_{\text{SSW}}$ and $\widehat \tau_{\text{PSW}}$ tend to perform similarly and well, where there may be a slight advantage to the latter estimator. This empirically shows that $\widehat \tau_{\text{SSW}}$ and $\widehat \tau_{\text{PSW}}$ are relatively robust to certain departures from their requisite identification assumptions i.e., when the identification assumptions underlying the other estimator hold. Performance instead appears to be most affected by modeling choice, where we see analogous patterns for the SSW and PSW estimators.

The results demonstrate that there is usually benefit relative to bias for explicitly accounting for clustering effects in modeling the mortality event via a GLMM as opposed to a GLM. The differences in their bias are relatively negligible in most scenarios, but the bias of $\widetilde \tau_{\text{SSW}}$ and $\widetilde \tau_{\text{PSW}}$ becomes more pronounced when the treatment effect on survival status further deviates from zero. In general, the fixed effects parameters in marginal logistic models tend to be attenuated towards zero,\citep{zeger1988models} which accords with the opposite sign bias we see with the GLM model when there are larger treatment effects. 
For the most part, when there are small treatment effects, the coverage rates are comparable across these modeling techniques, and they all are near to the nominal $95\%$ rate. However, when the value of $\lambda$ is higher, such as 0.3 (and occasionally 0.15), the constructed confidence intervals for $\widehat \tau_{\text{SSW}}$ and $\widehat \tau_{\text{PSW}}$ tend to suffer from undercoverage as compared $\widetilde \tau_{\text{SSW}}$ and $\widetilde \tau_{\text{PSW}}$ even though the variance estimates of the former are higher on average (regardless of small-sample corrections). Discrepancies here might be explained by the inferential trade-offs of using a marginal model in lieu of a conditional model in the presence of  certain values of ICC.\citep{murray2004design,rodriguez1995assessment,turner2017review2,thompson2022cluster} 
Given our observations in the small treatment effect scenarios, opting for $\widetilde \tau_{\text{SSW}}$ and $\widetilde \tau_{\text{PSW}}$ may suffice even if they cost some in terms of bias. However, the story for inference is different when the treatment effect on survival is large, particularly for higher values of $\lambda$. The interval estimates for $\widehat \tau_{\text{SSW}}$ and $\widehat \tau_{\text{PSW}}$ achieve closer to nominal coverage as opposed to $\widetilde \tau_{\text{SSW}}$ and $\widetilde \tau_{\text{PSW}}$, which demonstrate noticeable undercoverage. In our simulations, all of the average estimated variances are near to their empirical counterparts signifying that our variance estimators perform adequately in finite samples and bias, particularly for the GLM, is mostly driving the coverage results. The results for additional treatment effects and survival ICC are shown in Tables S1-S3 of Supplemental Materials B and in general are consistent with those presented above. In selected rerun simulations, we have also examined the performance of the cluster bootstrap for generating variance estimates and confidence intervals, presented in Table S4 of Supplemental Materials B. We observe that the results are similar (if not slightly worse) to the proposed variance estimators, but the computation time required by our variance estimators is on average reduced by more than ten-fold for GLMM modeling and more than sixty-fold for GLM modeling relative to the corresponding non-parametric bootstrap methods with 250 replicates.

\section{Application to the RESTORE Trial}\label{sec:applications}

We apply the weighting methods to estimate SACE in the Randomized Evaluation of Sedation Titration for Respiratory Failure (RESTORE) trial, a parallel-arm CRT investigating the effect of protocolized sedation on mechanical ventilation among children with acute respiratory failure, first analyzed by Curley et al.\citep{curley2015protocolized} In this study, $n_c=31$ pediatric care units (PICUs) were randomized to receive a goal-directed nurse-implemented sedation protocol to treat their mechanically ventilated patients. Among them, 17 PICUs were assigned to the active intervention and 14 others acted as the control where they were administered the status quo treatment. The cluster sizes range from 12 to 272 with a median size of 57, and there are a total of $n=$2449 children with 1225 in intervention and 1224 in control. There are 110 (4.5$\%$) observed deaths of which 47 (3.8$\%$) are in the intervention group and 63 (5.1$\%$) are in the control group.

In our analysis of RESTORE, the outcome is duration on a mechanical ventilator during a 28-day period. Patients who do not reach their follow-up period due to ICU-death are considered to have their outcome truncated, which justifies targeting the SACE estimand. We apply our weighting estimators of SACE to the RESTORE data, where the survival status models adjust for variables suggested in the initial analysis.\citep{curley2015protocolized} The covariates are age in years, Pediatric Risk of Mortality (PRISM) III-12 score (on log-scale), an aggregated measure of physiological status,\citep{pollack2016pediatric} and Pediatric Overall Performance Category (POPC) $>$ 1, an assessment of functional morbidity and cognitive impairment, \citep{fiser1992assessing} and for the GLMM, there is a random intercept for PICU facility.

The SACE estimates where survival status is modeled using a mixed effects logistic regression model are $\widehat \tau_{\text{SSW}}=-0.152(-1.657,1.353)$ and $\widehat \tau_{\text{PSW}}=-0.152 (-1.652,1.349)$. The SACE estimates where survival status is modeled using a logistic regression model are similar $\widetilde \tau_{\text{SSW}}=-0.148(-1.444,1.148)$ and $\widetilde\tau_{\text{PSW}}=-0.151(-1.442,1.141)$. All standard errors are computed using the proposed sandwich variance estimators with a degrees of freedom correction. As displayed in Table~\ref{tab:appliedres}, differences between the sandwich variance estimates (uncorrected) are negligible, which is expected because the estimated ICC for survival is small, only $0.026$. 

The negative estimates suggest that the protocolized sedation reduces duration on mechanical ventilation among always-survivors on average, but this effect is not statistically significant at the 0.05 level. Additionally, the nearly identical results for SSW and PSW estimators illustrate their robustness relative to the two mutually exclusive identification assumptions that underlie these weighting estimators. For example, it may be such that survival monotonicity is not valid here because the \textit{a priori} assertion that more enhanced sedation practices reduce or maintain survival rates is dubious. Nonetheless, the PSW estimates mirror those of SSW.
 
\begin{table}
    \centering
    \captionsetup{labelfont=bf,justification=raggedright,singlelinecheck=false}
    \caption{Treatment effect estimates of sedation protocol on mechanical ventilation duration at 28 days in RESTORE data. The proposed SACE and variance estimates based on differing assumptions and modeling techniques are provided in Subtable (a). Coefficients, $\widehat{\beta}_{\text{trt}}$, and variance estimates from standard multilevel models fit on patients who were observed to survive are included in Subtable (b). No degrees of freedom adjustments are made.}
    \begin{subtable}{0.48\linewidth}
        \centering
        \caption{SACE Estimators}
        \begin{tabular}{cc}
            \toprule
            Name & Estimate $(\widehat{\text{Var}})$ \\
            \midrule
            $\widehat{\tau}_{\text{SSW}}$ & -0.152 (0.437) \\
            $\widehat{\tau}_{\text{PSW}}$ & -0.152 (0.435) \\
            $\widetilde{\tau}_{\text{SSW}}$ & -0.148 (0.437) \\
            $\widetilde{\tau}_{\text{PSW}}$ & -0.151 (0.434) \\
            \bottomrule
        \end{tabular}
        \label{subtab:sace}
    \end{subtable}
    \hfill
    \begin{subtable}{0.48\linewidth}
        \centering
        \caption{Coeff. Estimators, $\widehat{\beta}_{\text{trt}}$, among Observed Survivors}
        \begin{tabular}{lc}
            \toprule
            Model & Estimate $(\widehat{\text{Var}})$ \\
            \midrule
            Unadjusted linear GEE & -0.149 (0.435) \\
            Unadjusted LMM & -0.465 (0.422) \\
            Adjusted linear GEE & -0.103 (0.399) \\
            Adjusted LMM & -0.440 (0.428) \\
            \bottomrule
        \end{tabular}
        \label{subtab:coeff}
    \end{subtable}
    \label{tab:appliedres}
\end{table}

As a further illustration, we fit unadjusted and adjusted standard multilevel models to the RESTORE data removing patients whose death was reported at 28 days. The models are a GEE with Gaussian-type mean and variance functions with an independence working correlation, also known as linear regression with cluster-robust standard errors, and a linear mixed model (LMM) with a random intercept for PICU. We considered these models as they are among the commonly used regression methods for analyzing CRTs but may not produce causal estimates in the presence of truncation by death. In Table~\ref{tab:appliedres}, the non-mortal treatment effect estimate, which is the coefficient of treatment $\widehat \beta_{\text{trt}}$ in the models, is reported alongside the variance estimate, $\widehat{\text{Var}}$ $ (\widehat \beta_{\text{trt}})$. These values are juxtaposed with the estimates for SACE (no degrees of freedom adjustments are made). We observe that the results for the unadjusted GEE are similar to those that we have with the SACE estimators. However, $\widehat \beta_{\text{trt}}$ does not target a clearly interpretable causal estimand because conditioning on survival breaks the balance of confounders across treatment groups and can lead to comparisons defined on different subpopulations. The similarity in estimates alone does not support the general application of such methods due to the ambiguity in the target estimand and may likely be attributed to the relatively low mortality rates in the RESTORE trial. Moreover, we see that the estimates for the remaining models, in particular the LMM models, are not comparable to those that we obtained for our SACE estimators. These differences may in fact be related to the selection bias introduced by conditioning on survivors and any distributional assumptions about the observed outcome e.g. the LMM assumes the non-mortal outcome is normally distributed conditional on covariates and the random intercept. In contrast, the weighting methods avoid the need to consider outcome models that are often difficult to correctly specify, and they only require selecting a binary regression model to capture the likelihood of survival until the time of outcome measurement. 

\section{Discussion}\label{sec:discussion}

For randomized trials targeting the effect of a binary treatment on a non-mortal outcome, participants who do not survive to their follow-up visit will have truncated outcome measurements. Estimating effects conditional on realizations of post-treatment variables such as survival status is a well-known potential source of selection bias that depends on underlying causal pathways. Moreover, when clusters are the randomization unit as in CRTs, there is an additional source of variation in patient outcomes due to within-group correlation. For this multifaceted problem, we target the SACE, a treatment effect conditional on individuals who would survive under either treatment received, and we discuss two sets of assumptions that allow for point identification of the SACE, notably accounting for unobservable sources of variation via a group-level latent variable. We develop two corresponding estimators for the SACE, denoted by SSW and PSW, based on the theory of M-estimation, where their asymptotic distributions have sandwich variance expressions enabling cluster-robust variance estimation.

Results of the simulation studies, which vary both the magnitudes of the within-cluster correlation and the treatment effect for survival status, suggest that the SACE estimators perform well in terms of empirical bias, and their confidence intervals generally achieve nominal coverage even for a small to modest number of clusters. There are occasional issues in undercoverage, but sources of this phenomenon have been documented in the past and may simply be remedied by other choices of small-sample corrections. Critically, both the SSW and PSW estimators prove to be robust to violations to one of their identifying assumptions -- conditional survival independence for SSW and survival monotonocity for PSW. Furthermore, for these estimators, we consider models to predict survival status that do and do not explicitly address within-cluster correlation :  mixed effects logistic regression (a type of GLMM) and logistic regression (a type of GLM) respectively. Performance in simulations across these modeling decisions is similar. Nevertheless, from the standpoint of addressing latent cluster-induced variation, the GLMM is a more principled approach compared to the GLM, and the simulation study shows that the empirical bias for the SACE estimator is in general smaller when the former is used. But, we demonstrate that in finite samples where the ICC for survival is also relatively low, the choice to use a GLMM may not be as stable (due in part to numerical methods), and one may prefer the GLM with a cluster-robust variance expression for inference, except if there is a previously known strong effect of the treatment on survival. The above observations bear out in our application to the RESTORE data, where there are negligible differences in SACE interval estimates across the assumption sets and modeling choices. However, the SACE estimates may differ from estimation of treatment effects among observed survivors, which have no causal interpretation, and the  magnitude of this difference may be amplified by (incorrect) assumptions when modeling the outcome distribution. To enable users to implement our methods, we provide an \texttt{R} package\citep{isenberg2024sace} for computing SACE interval estimates based on choices of the set of assumptions and survival status model. While we have a non-parametric bootstrap option in this package, the default estimation of asymptotic variances is computationally more efficient than bootstrapping and avoids complications of model fitting on bootstrapped samples when there is a small number of clusters. 

The limitations and future considerations for our work relate to assumptions on and utilization of the data. In the RESTORE study, we analyze the data by focusing on cross-sectional patient information with respect to their duration on a mechanical ventilator after 28 days, potentially not leveraging all the information from these data. We do not perform a time-to-event analysis with respect to extubation as done in the original study.\citep{curley2015protocolized} One immediate reason for this choice is that most standard survival models, such as the Cox regression fit initially to the RESTORE data, target the hazard ratio (HR), which is a parameter that conditions on survival up to time $t$, a post-treatment variable. Therefore, if survivor bias is present at time $t$, the HR inherently cannot be identified with any causal quantity for treatment effect.\citep{aalen2015does,martinussen2020subtleties,axelrod2023sensitivity} Moreover, in the case of non-mortal outcomes in time-to-event studies like RESTORE, determining the target causal estimand is even more complex due to the survivor bias associated with the non-mortal event and the HR as well as the fact that death disrupts survival time. Since death is a terminal event that precludes full measurement of the non-mortality outcome but not vice-versa, these events are often referred to as semicompeting risks.\citep{fine2001semi} Xu et al.\citep{xu2022bayesian}  propose a continuous time version of SACE expanding on existing methods for analyzing semicompeting risks to tackle this problem. They employ a Bayesian nonparametric approach to estimate this SACE quantity for individual randomized trials. To our knowledge, no such continuous time SACE has been applied to CRTs. Further developments for SACE are required to address the complexities introduced by clustering and time-to-event analyses.  

Moreover, the proposed estimation of the SACE estimands relies on cross-world identification assumptions i.e., worlds defined by intervening on $a=1$ and $a=0$. This is a debated framework in causal inference because there is no setting, even hypothetically, where simultaneous interventions could occur and assumptions across them are able to be verified.\citep{richardson2013single} As an alternative framework, Stensrud et al.\citep{stensrud2022conditional} propose estimands based on the notion of separable effects which avoid these cross-world assumptions and can be applied to the truncation by death problem. But, separable effects requires a significant conceptual exercise itself -- one of decomposing treatment disjointly into its effect on survival and its effect on the non-mortal outcome. Despite the inherent untestability of cross-world assumptions, we opt for them because we feel that these assumptions as well as the SACE estimand that they identify are more easily translated to language accessible to practitioners. Moreover, they can be generalized to other intercurrent events,\citep{ding2017principal} where the developments for SACE in this article can be extended accordingly. We have also assumed that there is no (marginal) dependence between the individual-level survival and non-mortal outcomes and cluster size (i.e., no informative cluster size), which can impact the validity of the proposed estimators for targeting individual-level effects.\citep{kahan2023estimands} When informative cluster size is present, group-level and individual-level estimands may not overlap, necessitating a more principled approach to defining identification assumptions and compatible estimators as others have ventured at in recent literature using likelihood based approaches.\citep{bugni_inference_2023,wang2024model} Another fundamental assumption is related to the mechanism of randomization, where we suppose that our data come from parallel-arm CRTs. However, a comprehensive review of analyses of CRTs in critical care units shows that less than half of them are based on true parallel-arm trials with the remaining being crossover or stepped wedge.\citep{cook2021rationale} The emergence of these types of randomized trials, motivated by practical, ethical, and statistical reasons, \citep{crespi2016improved,hemming2015stepped} warrants an exploration of estimands for SACE that can accommodate different randomization designs.

\section*{Acknowledgments}
Research in this article was supported by the Patient-Centered Outcomes Research Institute\textsuperscript{\textregistered} (PCORI\textsuperscript{\textregistered} Award ME-2020C1-19220), and the United States National Institutes of Health (NIH), National Heart, Lung, and Blood Institute (grant number R01-HL168202). All statements in this report, including its findings and conclusions, are solely those of the authors and do not necessarily represent the views of the NIH or PCORI\textsuperscript{\textregistered} or its Board of Governors or Methodology Committee. The data example was prepared using the RESTORE Research Materials obtained from the National Heart, Lung, and Blood Institute (NHLBI) Biologic Specimen and Data Repository Information Coordinating Center (BioLINCC) and does not necessarily reflect the opinions or views of the RESTORE study or the NHLBI.

\bibliographystyle{unsrtnat}
\bibliography{Bibliography.bib}

\appendix

\section*{Supplemental Materials: Weighting methods for truncation by death in cluster-randomized trials}

\section{Proofs, Technical Details, and Extensions}\label{sec:appendix1}

Recall that $p_{ij}^{a}(X_i,C_i):=P(S_{ij}=1|A_i=a,X_i,C_i)=E(S_{ij}|A_i=a,X_i,C_i)$. For $a=0,1$ by A2 (randomization and positivity) and cluster-level SUTVA, we have the identification:
\begin{equation}P(S_{ij}(a)=1|X_i,C_i)=P(S_{ij}(a)=1|A_i=a,X_i,C_i)=P(S_{ij}=1|A_i=a,X_i,C_i) \end{equation}
For example, one working model we consider for $P(S_{ij}=1|A_i,X_i,C_i)$ is a mixed effects logistic regression model (GLMM) with a fixed effect parameter for treatment such that $p_{ij}^{a}(X_i,C_i)$ is obtained by setting $A_i=a$.

\subsection{Proof of Identity 1}\label{sec:app11}

This proof holds for $\mu(1)$ and $\mu(0)$ since $a$ is arbitrary. By law of iterated expectations (LIE)
\begin{equation}E \left \{ Y_{ij}(a)S_{ij}(a)S_{ij}(1-a) \right \}=E \left \{ E \left (Y_{ij}(a)S_{ij}(a)S_{ij}(1-a)|X_i,C_i \right) \right \}\end{equation}
and
\begin{equation}E \left \{S_{ij}(a)S_{ij}(1-a)\right \} = E \left \{ E \left (S_{ij}(a)S_{ij}(1-a)|X_i,C_i \right) \right \}\end{equation}
We show by S1A4 (conditional survival independence), and S1A5 (strong partial principal ignorability), we have 
\begin{equation}\label{eq:nuexp}E \left \{Y_{ij}(a)S_{ij}(a)S_{ij}(1-a)|X_i,C_i \right \}=E\{Y_{ij}(a)S_{ij}(a)E\{S_{ij}(1-a)|X_i,C_i\}|X_i,C_i\}\end{equation}
and by S1A4
\begin{equation} \label{eq:xiexp} E \left \{S_{ij}(a)S_{ij}(1-a)|X_i,C_i \right \}=E\{S_{ij}(a)E\{S_{ij}(1-a)|X_i,C_i\}|X_i,C_i\}\end{equation}
For the proof of Equation~\eqref{eq:nuexp}.
\begin{align}
E& \left \{Y_{ij}(a)S_{ij}(a)S_{ij}(1-a)|X_i,C_i \right \}  
\nonumber \\=&E \left [ E \left\{Y_{ij}(a)S_{ij}(a)S_{ij}(1-a)|S_{ij}(a),S_{ij}(1-a),X_i,C_i \right \}|X_i,C_i \right ] \hspace{.3cm} \text{(by LIE)} \\
=&E \left\{Y_{ij}(a)|S_{ij}(a)=1,S_{ij}(1-a)=1,X_i,C_i \right \}P(S_{ij}(a)=1,S_{ij}(1-a)=1|X_i,C_i) \nonumber \\
=&E \left\{Y_{ij}(a)|S_{ij}(a)=1,X_i,C_i \right \}P(S_{ij}(1-a)=1|X_i,C_i)P(S_{ij}(a)=1|X_i,C_i) \hspace{.3cm} \text{(by S1A4\&A5)} \nonumber \\
=&E \left [ E\left \{ Y_{ij}(a)S_{ij}(a)E(S_{ij}(1-a)|X_i,C_i)|S_{ij}(a),X_i,C_i \right \}|X_i,C_i \right] \nonumber \\
=& E\{Y_{ij}(a)S_{ij}(a)E(S_{ij}(1-a)|X_i,C_i)|X_i,C_i\}  \hspace{.3cm} \text{(by LIE)} \nonumber
\end{align}

Note that A3 is implicitly used to represent the initial conditional expectation in terms of the individual-level distributions, making S1A4 and S1A5 usable. Equation~\eqref{eq:xiexp} will be left without proof as it's immediate from the assumptions. From \eqref{eq:nuexp} and \eqref{eq:xiexp}:
\begin{align}
\begin{split}
\mu(a)&=\frac{E \left\{E(Y_{ij}(a)S_{ij}(a)p^{1-a}_{ij}(X_i,C_i)|X_i,C_i)\right\}}{E\left\{E(S_{ij}(a)p^{1-a}_{ij}(X_{i},C_i)|X_i,C_i)\right\}}\\
&=\frac{E \left\{E(Y_{ij}(a)S_{ij}(a)p^{1-a}_{ij}(X_i,C_i)|A_i=a,X_i,C_i)\right\}}{E\left\{E(S_{ij}(a)p^{1-a}_{ij}(X_{i},C_i)|A_i=a,X_i,C_i)\right\}} \hspace{.3cm} \text{(by A2)}\\
&=\frac{E \left\{E(Y_{ij}S_{ij}p^{1-a}_{ij}(X_i,C_i)|A_i=a,X_i,C_i)\right\}}{E\left\{E(S_{ij}p^{1-a}_{ij}(X_i,C_i)|A_i=a,X_i,C_i)\right\}} \hspace{.3cm} \text{(by cluster-level SUTVA)}\\
&=\frac{E \left\{Y_{ij}I(A_i=a)S_{ij}p^{1-a}_{ij}(X_i,C_i)\right\}}{E\left\{I(A_i=a)S_{ij}p^{1-a}_{ij}(X_i,C_i)\right\}} \hspace{.3cm} \text{(by A2 and LIE)}
\end{split}
\end{align}
where non-trivial survival prevents division by 0. 

\subsection{Proof of Identity 2}\label{sec:app12}
Let's start with identification of $\mu(0)$. Recall that S2A4 is survival monotonicity $S_{ij}(1) \ge S_{ij}(0)$ a.s. We have
\begin{align}
\begin{split}
\mu(0)&=\frac{E \{Y_{ij}(0)S_{ij}(0) \}}{E \{S_{ij}(0)\}} \hspace{.3cm} \text{(by S2A4)}\\
&=\frac{E \{Y_{ij}(0)S_{ij}(0)|A_i=0 \}}{E \{S_{ij}(0)|A_i=0\}} \hspace{.3cm} \text{(by A2)}\\
&=\frac{E \{Y_{ij}S_{ij}|A_i=0 \}}{E \{S_{ij}|A_i=0\}}=\frac{E \{Y_{ij}(1-A_i)S_{ij} \}}{E \{(1-A_i)S_{ij}\}}  \hspace{.3cm} \text{(by cluster-level SUTVA and A2)}
\end{split}
\end{align}
Now, let's consider $\mu(1)$, which also requires partial principal ignorability, S2A5.
\begin{align}
 \mu(1) &= \frac{E\left \{Y_{ij}(1)S_{ij}(1)S_{ij}(0) \right \}}{E\left \{S_{ij}(0) \right \}} \hspace{.2cm} \text{(by S2A4)}\\
 &\label{eq:lie}=\frac{E\left \{E(Y_{ij}(1)|S_{ij}(1)=1,S_{ij}(0)=1,X_i,C_i)P(S_{ij}(1)=1,S_{ij}(0)=1|X_i,C_i) \right \}}{E \left \{E(S_{ij}(0)|X_i,C_i) \right \}}\\
 & =\frac{E\left \{E(Y_{ij}(1)|S_{ij}(1)=1,S_{ij}(0)=1,X_i,C_i)E(S_{ij}(0)|X_i,C_i) \right \}}{E\left \{E(S_{ij}(0)|X_i,C_i) \right \}} \hspace{.3cm} \text{(by S2A4)} \nonumber \\
&\label{eq:idu1s2}=\frac{E\left \{ E(Y_{ij}(1)p^{0}_{ij}(X_i,C_i)|A_i=1,S_{ij}(1)=1,X_i,C_i) \right \}}{E\left \{E(S_{ij}(0)|A_i=1,X_i,C_i) \right \}} \hspace{.3cm} \text{(by S2A5 and A2)} \nonumber \\
&=\frac{E\left \{ E\left[\frac{Y_{ij}(1)A_iS_{ij}(1)p^{0}_{ij}(X_i,C_i)}{P(S_{ij}(1)=1|A_i=1,X_i,C_i)P(A_i=1|X_i,C_i)} \bigg \vert X_i,C_i \right] \right \}}{E\left \{E\left(\frac{A_iS_{ij}(0)}{P(A_i=1|X_i,C_i)} \bigg \vert X_i,C_i \right )\right\}}\hspace{.3cm} \text{(by A2, non-trivial survival)} \nonumber \\
&=\frac{E\left \{ E\left[\frac{Y_{ij}A_iS_{ij}p^{0}_{ij}(X_i,C_i)}{p^{1}_{ij}(X_i,C_i)P(A_i=1)} \bigg \vert X_i,C_i \right] \right \}}{E\left \{E\left(\frac{A_iS_{ij}(0)}{P(A_i=1)} \bigg \vert X_i,C_i \right )\right\}} \hspace{.3cm} \text{(by cluster-level SUTVA and A2)} \nonumber \\
&=\frac{E\left \{ E\left[\frac{Y_{ij}A_iS_{ij}p^{0}_{ij}(X_i,C_i)}{p^{1}_{ij}(X_i,C_i)} \bigg \vert X_i,C_i \right] \right \}}{E\left \{A_i E(S_{ij}(0)| X_i,C_i)\right\}} \hspace{.3cm} \text{(by A2)} \nonumber \\
&=\frac{E\left \{ E\left[\frac{Y_{ij}A_iS_{ij}p^{0}_{ij}(X_i,C_i)}{p^{1}_{ij}(X_i,C_i)} \bigg \vert X_i,C_i \right] \right \}}{E\left \{A_i p^{0}_{ij}(X_i,C_i)\frac{E(S_{ij}(1)|X_i,C_i)}{E(S_{ij}(1)|X_i,C_i)}\right\}} \nonumber \\
&=\frac{E\left \{ E\left[\frac{Y_{ij}A_iS_{ij}p^{0}_{ij}(X_i,C_i)}{p^{1}_{ij}(X_i,C_i)} \bigg \vert X_i,C_i \right] \right \}}{E\left \{A_i p^{0}_{ij}(X_i,C_i)\frac{E(S_{ij}(1)|A_i,X_i,C_i)}{E(S_{ij}(1)|X_i,C_i)}\right\}} \hspace{.3cm} \text{(by A2)} \\
&=\frac{E\left \{ E\left[\frac{Y_{ij}A_iS_{ij}p^{0}_{ij}(X_i,C_i)}{p^{1}_{ij}(X_i,C_i)} \bigg \vert X_i,C_i \right] \right \}}{E\left \{ E \left(\frac{A_iS_{ij}(1)p^{0}_{ij}(X_i,C_i)}{E(S_{ij}(1)|X_i,C_i)} \bigg \vert A_i,X_i,C_i \right)\right\}} \nonumber \\
&=\frac{E\left \{\frac{Y_{ij}A_iS_{ij}p^{0}_{ij}(X_i,C_i)}{p^{1}_{ij}(X_i,C_i)} \right \}}{E\left \{\frac{A_iS_{ij}p^{0}_{ij}(X_i,C_i)}{p^{1}_{ij}(X_i,C_i)}\right\}} \hspace{.3cm} \text{(by LIE and cluster-level SUTVA)} \nonumber
\end{align}
where in \eqref{eq:lie} we use LIE exactly as written out in detail in Identity 1. Note again that A3 is implicitly used to represent the initial conditional expectation in terms of the individual-level distributions, making S2A4 and S2A5 usable. 

\subsection{Sketch of Asymptotic Arguments for M-Estimation}\label{sec:app13}

We elaborate on the asymptotic arguments in Section 2 of the main paper. Recall that the observed vector of fixed effects $D_{ij}$ and $\beta$ have dimension $p$. We consider the values of parameter $\theta^T=(\beta^T,\sigma^2_b,\mu(1),\mu(0))$ of dimension $(p+3) \times 1$. 
For each $i$, we denote $m(D_i,S_i,Y_i; \theta)=m_{i}(\theta)$, which are mutually independent due to cluster randomization ($n_i <\infty$). These arguments are general, so for simplicity of notation, we only use the second subscript as in $m_{(\cdot),i}$ when necessary to distinguish the SSW and PSW estimators (after Equation~\eqref{eq:dldsig}). If $E_\theta\{m_{i}(\theta)\}=0$ for any $\theta$, we can construct unbiased estimating equations of dimension $(p+3) \times 1$
\begin{equation}\frac{1}{n_c}\sum_{i=1}^{n_c}m_{i}(\theta)=0\end{equation}
whose solution is the M-estimator $\widehat \theta$ . Following the general set up of Stefanski and Boos\citep{stefanski2002calculus}, we have the first-order Taylor polynomial approximation centered at the true parameter $\theta=\theta_0$:
\begin{equation}\label{eq:taylor}0=\frac{1}{\sqrt{n_c}} \sum_{i=1}^{n_c}m_i(\widehat \theta)=\frac{1}{\sqrt{n_c}} \sum_{i=1}^{n_c} m_i( \theta_0)+\frac{1}{n_c}\sum_{i=1}^{n_c}\frac{\partial m_i(\theta_0)}{\partial \theta^T}\sqrt{n_c}(\widehat \theta-\theta_0)+\sqrt{n_c}R_{n_c}\end{equation}
If the following expression is non-singular:
\begin{equation}\frac{1}{n_c}\sum_{i=1}^{n_c} \frac{\partial m_i(\theta_0)}{\partial \theta^T}\end{equation}
then we can rewrite Equation~\eqref{eq:taylor} as:
\begin{equation} \label{eq:taylorrearr} \sqrt{n_c}(\widehat \theta -\theta_0) = -\left [ \frac{1}{n_c}\sum_{i=1}^{n_c} \frac{\partial m_i(\theta_0)}{\partial \theta^T} \right]^{-1}\frac{1}{\sqrt{n_c}}\sum_{i=1}^{n_c}m_i(\theta_0)+\sqrt{n_c}R^{*}_{n_c}\end{equation}

If under the sufficient regularity and data conditions, including for example Lindeberg's condition for triangular arrays (or the stronger but more accessible Lyapunov's condition) for independent but not necessarily identically distributed data, we have that 
\begin{equation}\frac{1}{n_c}\sum_{i=1}^{n_c} \frac{\partial m_i(\theta_0)}{\partial \theta^T} \xrightarrow{p} \lim_{n_c \rightarrow \infty} \frac{1}{n_c} \sum_{i=1}^{n_c} E \left \{\frac{\partial m_{i}(\theta_0)}{\partial \theta^T} \right \}\end{equation} 
\begin{equation}\sqrt{n_c} \frac{1}{n_c} \sum_{i=1}^{n_c}m_i(\theta_0) \xrightarrow{d} \text{MVN}\left(0,\lim_{n_c \rightarrow \infty} \frac{1}{n_c} \sum_{i=1}^{n_c} \text{Var}\{m_i(\theta_0)\}\right)\end{equation}
and conditions for the remainder to vanish \citep{serfling2009approximation}
\begin{equation}\sqrt{n_c}R^{*}_{n_c} = o_p(1) \end{equation}
The continuous mapping and Slutsky's theorems would ensure that 
\begin{equation}\sqrt{n_c}(\widehat \theta - \theta_0) \xrightarrow{d} \text{MVN} \left(0, V_{\theta_0} \right)\end{equation} where
\begin{align}\label{eq:truev}
\begin{split}
V_{\theta_0}=&\left[\lim_{n_c \rightarrow \infty} \frac{1}{n_c}\sum_{i=1}^{n_c}E \left \{\frac{\partial m_{i}(\theta_0)}{\partial \theta^T} \right \} \right]^{-1} \lim_{n_c \rightarrow \infty} \frac{1}{n_c} \sum_{i=1}^{n_c} \text{Var}\{m_i (\theta_0)\}  \\
&\times \left[\lim_{n_c \rightarrow \infty} \frac{1}{n_c}\sum_{i=1}^{n_c}E \left \{\frac{\partial m_{i}(\theta_0)}{\partial \theta^T} \right \} \right]^{-T}
\end{split}
\end{align}
with $\text{Var}\{m_{i}(\theta_0)\}=E\{m_{i}(\theta_0)m_{i}(\theta_0)^T\}$ for all $i$. For consistency results and corresponding necessary regularity conditions for replacing parameters in the sandwich variance expression with their estimators as in $\widehat V_{\widehat \theta}$ consult Iverson and Randles\citep{iverson1989effects}. Recall $k=(0^T,0,1,-1)^T$ and we wish to target $k^T \theta_0$. We consider estimators $k^T \widehat \theta = \widehat \mu(1) - \widehat \mu(0)$.
By the delta method (i.e., a first order Taylor approximation of a composition of functions as in Equation~\eqref{eq:taylor}), $\sqrt{n_c}(k^T \widehat \theta-k^T \theta_0) \xrightarrow{d} N(0,k^TV_{\theta_0}k)$
since $\nabla_{\theta} k^T \theta = k$. Moreover, under the above conditions\citep{iverson1989effects} and the continuous mapping theorem, $k^T \widehat V_{\widehat \theta}k$ will converge in probability to $k^TV_{\theta_0}k$, where

\begin{equation}
\widehat V_{\widehat \theta}=\left[\frac{1}{n_c}\sum_{i=1}^{n_c} \left \{\frac{\partial m_{i}(\widehat \theta)}{\partial \theta^T}\right \} \right]^{-1} \frac{1}{n_c}\sum_{i=1}^{n_c}m_{i}(\widehat \theta)m_{i}(\widehat \theta)^T  \left[\frac{1}{n_c} \sum_{i=1}^{n_c} \left \{\frac{\partial m_{i}(\widehat \theta)}{\partial \theta^T} \right \}\right]^{-T}
\end{equation}
An approximate of the variance of the SACE estimators is then given by $(k^T \widehat V_{\widehat \theta}k)/n_c$.

\subsection{Components of Estimating Functions and Matrices for Variance Expressions}\label{sec:app14}

Recall from Section 2.3 of the main paper, we need to stack unbiased estimating equations for which $\widehat \theta=(\widehat \beta^T,\widehat \sigma^2_b,\widehat \mu(1),\widehat \mu(0))^T$ is the solution. We write out the statement more completely. Integral expressions for vectors (and matrices) denote component wise integration.

$\widehat \beta$ and $\widehat \sigma^2_b$ are the values which maximize the following ``observed-data" likelihood:
\begin{align}
L(S|D,\beta, \sigma^2_b)&=\int^{\infty}_{-\infty}  L(S|D,b, \beta, \sigma^2_b)L(b|\sigma^2_b)db\\
&=\prod_{i=1}^{n_c} \int^{\infty}_{-\infty}   L_i(S_i|D_i,b_i, \beta, \sigma^2_b)L(b_i|\sigma^2_b)db_i \nonumber \\
&= \prod_{i=1}^{n_c} \int^{\infty}_{-\infty}   \left[\prod_{j=1}^{n_i} \frac{[\exp(D_{ij}^T \beta + b_i)]^{S_{ij}}}{1+\exp(D_{ij}^T \beta + b_i)}\right]\frac{1}{\sqrt{2\pi\sigma^2_b}}\exp \left(-\frac{b_i^2}{2\sigma^2_b}\right)db_i \nonumber \\
& \propto (\sigma^2_b)^{-\frac{n_c}{2}}\prod_{i,j}[\exp(D_{ij}^T \beta)]^{S_{ij}}  \prod_{i=1}^{n_c} \int^{\infty}_{-\infty} \left[\prod_{j=1}^{n_i} \frac{[\exp(b_i)]^{S_{ij}}}{1+\exp(D_{ij}^T \beta + b_i)}\right]\exp \left(-\frac{b_i^2}{2\sigma^2_b}\right)db_i \nonumber
\end{align}
Let 
\begin{equation} \label{eq:kernel} g_i(b_i)=\exp \left( \sum_{j=1}^{n_i} \left[S_{ij}b_i-\log \left(1+\exp(D_{ij}^T \beta + b_i) \right)\right]-\frac{b_i^2}{2\sigma^2_b}\right)\end{equation}
The log-likelihood is represented as (up to a constant)
\begin{align}l(S|D,\beta,\sigma^2_b)&=\sum_{i=1}^{n_c}l_i(S_i|D_i,\beta,\sigma^2_b)\\
&=\sum_{i=1}^{n_c} \left( -\frac{1}{2}\log(\sigma^2_b)+ \sum_{j=1}^{n_i}S_{ij} D_{ij}^T\beta +\log \left \{ \int^{\infty}_{-\infty} g_i(b_i) \right \}\right) \nonumber \end{align}
Assuming that you can commute the derivative and integral, you can obtain components,$\frac{\partial l(S|D,\beta,\sigma^2_b)}{\partial \beta}$ and $\frac{\partial l(S|D,\beta,\sigma^2_b)}{\partial \sigma^2_b}$, of the score vector whose expectation is zero for any $\gamma$. They are 
\begin{align} \label{eq:dldbeta}
\frac{\partial l(S|D,\beta,\sigma^2_b)}{\partial \beta}&=\sum_{i=1}^{n_c} \frac{\partial l_i(S_i|D_i,\beta,\sigma^2_b)}{\partial \beta} \\
&=\sum_{i=1}^{n_c} \left(\sum_{j=1}^{n_i}S_{ij}D_{ij} - \frac{ \int^{\infty}_{-\infty}  
g_i(b_i)\sum_{j=1}^{n_i}\frac{D_{ij}\exp(D_{ij}^T\beta+b_i)}{1+\exp(D_{ij}^T\beta+b_i)}db_i}{ \int^{\infty}_{-\infty}  
g_i(b_i)db_i}\right) \nonumber
\end{align}

\begin{equation} \label{eq:dldsig} \frac{\partial l(S|D,\beta,\sigma^2_b)}{\partial \sigma^2_b}=\sum_{i=1}^{n_c}\frac{\partial l_i(S_i|D_i,\beta,\sigma^2_b)}{\partial \sigma^2_b}=\sum_{i=1}^{n_c} \left(-\frac{1}{2\sigma^2_b}+\frac{1}{2\sigma_b^4}\frac{ \int^{\infty}_{-\infty}  
g_i(b_i)b_i^2db_i}{ \int^{\infty}_{-\infty}  
g_i(b_i)db_i}\right)\end{equation}
For any known $\gamma$ and $a=0,1$:
\begin{align}
\begin{split} 
E&_{\mu(a)}\left \{\sum_{j=1}^{n_i}Y_{ij}I(A_i=a)S_{ij}p^{1-a}
(X_i,C_i;\gamma) \right.
\\ & \left.-\mu(a) \sum_{j=1}^{n_i}I(A_i=a)S_{ij}p^{1-a}(X_i,C_i;\gamma)\right\}=0
\end{split}
\end{align}
for any $\mu(a)$ as defined with respect to observable data under Identity 1. Therefore, for our stacked estimating equations for SSW, $\sum_{i=1}^{n_c}m_{\text{SSW},i}(\theta)$, we have: 
\begin{equation} \label{eq:ms1i} m_{\text{SSW},i}(\theta)=\begin{pmatrix} \displaystyle \sum_{j=1}^{n_i}S_{ij}D_{ij} - \frac{ \int^{\infty}_{-\infty}  
g_i(b_i)\sum_{j=1}^{n_i}\frac{D_{ij}\exp(D_{ij}^T\beta+b_i)}{1+\exp(D_{ij}^T\beta+b_i)}db_i}{ \int^{\infty}_{-\infty}  
g_i(b_i)db_i} \\ 
\displaystyle-\frac{1}{2\sigma^2_b}+\frac{1}{2\sigma_b^4}\frac{ \int^{\infty}_{-\infty}  
g_i(b_i)b_i^2db_i}{ \int^{\infty}_{-\infty}  
g_i(b_i)db_i} \\  \sum_{j=1}^{n_i}Y_{ij}A_iS_{ij}p^0_{ij}(X_i,C_i;\gamma)-\mu(1)\sum_{j=1}^{n_i}A_iS_{ij}p^0_{ij}(X_i,C_i;\gamma)\\
 \sum_{j=1}^{n_i}Y_{ij}(1-A_i)S_{ij}p^1_{ij}(X_i,C_i;\gamma)-\mu(0)\sum_{j=1}^{n_i}(1-A_i)S_{ij}p^1_{ij}(X_i,C_i;\gamma)\end{pmatrix}\end{equation}
Similarly, for any known $\gamma$,
\begin{equation}E_{\mu(1)}\left (\sum_{j=1}^{n_i} \frac{Y_{ij}A_iS_{ij} p^{0}_{ij}(X_i,C_i;\gamma)}{ p^{1}_{ij}(X_i,C_i;\gamma)}-\mu(1)\sum_{j=1}^{n_i}   \frac{A_iS_{ij}p^{0}_{ij}(X_i,C_i;\gamma)}{p^{1}_{ij}(X_i,C_i;\gamma)}\right)=0\end{equation}
\begin{equation} E_{\mu(0)}\left (\sum_{j=1}^{n_i}Y_{ij}(1-A_i)S_{ij}-\mu(0)\sum_{j=1}^{n_i}(1-A_i)S_{ij} \right)=0 \end{equation}
for any $\mu(0)$ and $\mu(1)$ as identified by Identity 2. Then, for our stacked estimating equations for PSW, $\sum_{i=1}^{n_c}m_{\text{PSW},i}(\theta)$, we have:
\begin{equation}\label{eq:ms2i} m_{\text{PSW},i}(\theta)=\begin{pmatrix} \displaystyle \sum_{j=1}^{n_i}S_{ij}D_{ij} - \frac{ \int^{\infty}_{-\infty}  
g_i(b_i)\sum_{j=1}^{n_i}\frac{D_{ij}\exp(D_{ij}^T\beta+b_i)}{1+\exp(D_{ij}^T\beta+b_i)}db_i}{ \int^{\infty}_{-\infty}  
g_i(b_i)db_i} \\ 
\displaystyle-\frac{1}{2\sigma^2_b}+\frac{1}{2\sigma_b^4}\frac{ \int^{\infty}_{-\infty}  
g_i(b_i)b_i^2db_i}{ \int^{\infty}_{-\infty}  
g_i(b_i)db_i} \\ 
\sum_{j=1}^{n_{i}} \frac{Y_{ij}A_iS_{ij} p^{0}_{ij}(X_i,C_i;\gamma)}{ p^{1}_{ij}(X_i,C_i;\gamma)}-\mu(1)\sum_{j=1}^{n_{i}}   \frac{A_iS_{ij}p^{0}_{ij}(X_i,C_i;\gamma)}{p^{1}_{ij}(X_i,C_i;\gamma)}\\
 \sum_{j=1}^{n_i}Y_{ij}(1-A_i)S_{ij}-\mu(0)\sum_{j=1}^{n_i}(1-A_i)S_{ij}\end{pmatrix} \end{equation}

Recall we denote the middle matrices of the sandwich variance estimates with the following $M_{\text{SSW}}(\theta)=\sum_{i=1}^{n_c}m_{\text{SSW},i}(\theta)m_{\text{SSW},i}(\theta)^T$ and $M_{\text{PSW}}(\theta)=\sum_{i=1}^{n_c}m_{\text{PSW},i}(\theta)m_{\text{PSW},i}(\theta)^T$. We will denote the outer matrices by $B_{\text{SSW}}(\theta)=\sum_{i=i}^{n_c}\partial m_{\text{SSW},i}(\theta)/\partial \theta^T$ and $B_{\text{PSW}}(\theta)=\sum_{i=i}^{n_c}\partial m_{\text{PSW},i}(\theta)/\partial \theta^T$. The general form of the outer  matrices, $B_{(\cdot)}(\theta)$, is:
\begin{equation}
\begin{pmatrix}
\displaystyle \frac{\partial^2 l(S|D,\beta,\sigma^2_b)}{\partial \beta^T \partial \beta} & \displaystyle \frac{\partial^2 l(S|D,\beta,\sigma^2_b)}{\partial \sigma^2_b \partial \beta } & \displaystyle \frac{\partial^2 l(S|D,\beta,\sigma^2_b)}{\partial \mu(1) \partial \beta} & \displaystyle \frac{\partial^2 l(S|D,\beta,\sigma^2_b)}{\partial \mu(0) \partial \beta}\\
\displaystyle \frac{\partial^2 l(S|D,\beta,\sigma^2_b)}{\partial \beta^T \partial \sigma^2_b} &\displaystyle  \frac{\partial^2 l(S|D,\beta,\sigma^2_b)}{\partial (\sigma^2_b)^2} & \displaystyle  \frac{\partial^2 l(S|D,\beta,\sigma^2_b)}{\partial \mu(1) \partial \sigma^2_b} & \displaystyle  \frac{\partial^2 l(S|D,\beta,\sigma^2_b)}{\partial \mu(0) \partial \sigma^2_b}\\
\displaystyle \sum_{i=1}^{n_c} \frac{\partial m_{(\cdot),i,p+2}(\theta)}{\partial \beta^T} & \displaystyle \sum_{i=1}^{n_c} \frac{\partial m_{(\cdot),i,p+2}(\theta)}{\partial \sigma^2_b} & \displaystyle \sum_{i=1}^{n_c} \frac{\partial m_{(\cdot),i,p+2}(\theta)}{\partial \mu(1)} & \displaystyle \sum_{i=1}^{n_c}\frac{\partial m_{(\cdot),i,p+2}(\theta)}{\partial \mu(0)}\\
\displaystyle \sum_{i=1}^{n_c}\frac{\partial m_{(\cdot),i,p+3}(\theta)}{\partial \beta^T} & \displaystyle \sum_{i=1}^{n_c}\frac{\partial m_{(\cdot),i,p+3}(\theta)}{\partial \sigma^2_b} & \displaystyle \sum_{i=1}^{n_c}\frac{\partial m_{(\cdot),i,{p+3}}(\theta)}{\partial \mu(1) } & \displaystyle \sum_{i=1}^{n_c}\frac{\partial m_{(\cdot),i,{p+3}}(\theta)}{\partial \mu(0)}
\end{pmatrix}\end{equation}
where the third index of $m_{(\cdot),i,k}(\theta)$ represents the $k$-th component of the $i$-th vector function. We rewrite the matrix structures $B_{\text{SSW}}(\theta)$ and $B_{\text{PSW}}(\theta)$ for reference. For the SSW estimator, we have outer matrix:
\begin{equation}B_{\text{SSW}}(\theta)=\begin{pmatrix} 
B^{p \times p}_{11} & B^{p \times 1}_{12} & 0^{p \times 1} & 0^{p \times 1}\\
B^{1 \times p}_{21} &  B^{1 \times 1}_{2 2}& 0^{1 \times 1} & 0^{1 \times 1}\\
B^{1 \times p}_{\text{SSW},31} & 0^{1 \times 1} & B^{1 \times 1}_{\text{SSW},33}& 0^{1 \times 1}\\
B^{1 \times p}_{\text{SSW},41} & 0^{1 \times 1} & 0^{1 \times 1} & B^{1 \times 1}_{\text{SSW},44}
\end{pmatrix}\end{equation}
for the PSW estimator,
\begin{equation}
B_{\text{PSW}}(\theta)=\begin{pmatrix}
B^{p \times p}_{11} & B^{p \times 1}_{12} & 0^{p \times 1} & 0^{p \times 1}\\
 B^{1 \times p}_{21} &  B^{1 \times 1}_{2 2}& 0^{1 \times 1} & 0^{1 \times 1}\\
B^{1 \times p}_{\text{PSW},31} & 0^{1 \times 1} & B^{1 \times 1}_{\text{PSW},33}& 0^{1 \times 1}\\
0^{1 \times p} & 0^{1 \times 1} & 0^{1 \times 1} & B^{1 \times 1}_{\text{PSW},44}
\end{pmatrix}\end{equation}
where the superscripts denote the dimension of the submatrices ($\theta$ on the RHS is suppressed for notation simplicity). Therefore, we have estimates of the variances given by 
\begin{equation} \widehat{\text{Var}}(\widehat \tau_{(\cdot)}) \approx k^T B_{(\cdot)}(\widehat \theta_{(\cdot)})^{-1} M_{(\cdot)}(\widehat \theta_{(\cdot)}) B_{(\cdot)}(\widehat \theta_{(\cdot)})^{-T} k \end{equation}
For $B_{11}$ (we do not subscript as its common to both estimators) and using the chain rule, we have \begin{align} \label{eq:bmat}
\frac{\partial^2 l(S|D,\beta,\sigma^2_b)}{\partial \beta^T \partial \beta}=&\frac{\partial}{\partial \beta^T}\left\{-\sum_{i=1}^{n_c} \frac{ \int^{\infty}_{-\infty}  
g_i(b_i)\sum_{j=1}^{n_i} \frac{D_{ij}\exp(D_{ij}^T\beta+b_i)}{1+\exp(D_{ij}^T\beta+b_i)}db_i}{ \int^{\infty}_{-\infty}  
g_i(b_i)db_i}\right \}  \\
=&-\sum_{i=1}^{n_c}\left (\frac{ \int^{\infty}_{-\infty}  
g_i(b_i) \sum_{j=1}^{n_i} \frac{D_{ij}D_{ij}^T\exp(D_{ij}^T\beta+b_i)}{[1+\exp(D_{ij}^T\beta+b_i)]^2}db_i}{ \int^{\infty}_{-\infty}  
g_i(b_i)db_i}\right. \nonumber \\
&+\frac{ \int^{\infty}_{-\infty}  
g_i(b_i)\sum_{j=1}^{n_i}\frac{D_{ij}\exp(D_{ij}^T\beta+b_i)}{1+\exp(D_{ij}^T\beta+b_i)}\sum_{j'=1}^{n_i}\frac{D_{ij'}^T\exp(D_{ij'}^T\beta+b_i)}{1+\exp(D_{ij'}^T\beta+b_i)}db_i}{ \int^{\infty}_{-\infty}  
g_i(b_i)db_i} \nonumber \\
&\left. - \frac{\int^{\infty}_{-\infty}  
g_i(b_i)\sum_{j=1}^{n_i}\frac{D_{ij}\exp(D_{ij}^T\beta+b_i)}{1+\exp(D_{ij}^T\beta+b_i)}db_i}{\left[\int^{\infty}_{-\infty}  
g_i(b_i)db_i\right]^2} \right. \nonumber \\
& \left. \times \int^{\infty}_{-\infty}  
g_i(b_i)\sum_{j'=1}^{n_i}\frac{D_{ij'}^T\exp(D_{ij'}^T\beta+b_i)}{1+\exp(D_{ij'}^T\beta+b_i)}db_i \right) \nonumber 
\end{align}
For $B_{12}$, we have
\begin{align}
\frac{\partial^2 l(S|D,\beta,\sigma^2_b)}{\partial \sigma^2_b \partial \beta} &= \frac{\partial}{\partial \sigma^2_b}\left\{-\sum_{i=1}^{n_c} \frac{ \int^{\infty}_{-\infty}  
g_i(b_i)\sum_{j=1}^{n_i} \frac{D_{ij}\exp(D_{ij}^T\beta+b_i)}{1+\exp(D_{ij}^T\beta+b_i)}db_i}{ \int^{\infty}_{-\infty}  
g_i(b_i)db_i}\right \} \\
&=-\frac{1}{2\sigma_b^4}\sum_{i=1}^{n_c} \left( \frac{\int^{\infty}_{-\infty}  
g_i(b_i)\sum_{j=1}^{n_i} \frac{D_{ij}b_i^2\exp(D_{ij}^T\beta+b_i)}{1+\exp(D_{ij}^T\beta+b_i)}db_i}{\int^{\infty}_{-\infty}  
g_i(b_i)db_i}-\right. \nonumber \\
&\left. \frac{\int^{\infty}_{-\infty}  
g_i(b_i)\sum_{j=1}^{n_i} \frac{D_{ij}\exp(D_{ij}^T\beta+b_i)}{1+\exp(D_{ij}^T\beta+b_i)}db_i}{\left[\int^{\infty}_{-\infty}  
g_i(b_i)db_i\right]^2} \times \int^{\infty}_{-\infty}  
g_i(b_i)b_i^2db_i \right) \nonumber 
\end{align}
Indeed, by symmetry of the partial derivatives $B_{21}=B_{12}^T$. For $B_{22}$, we have
\begin{align}
\frac{\partial^2 l(S|D,\beta,\sigma^2_b)}{\partial \sigma^2_b \partial \sigma^2_b} &= \frac{\partial}{\partial \sigma^2_b} \left\{ -\frac{n_c}{2\sigma^2_b}+\frac{1}{2\sigma_b^4}\sum_{i=1}^{n_c}\frac{ \int^{\infty}_{-\infty}  
g_i(b_i)b_i^2db_i}{ \int^{\infty}_{-\infty}  
g_i(b_i)db_i} \right\}\\
=&\frac{n_c}{2\sigma^4_b}-\frac{1}{\sigma^6_b}\sum_{i=1}^{n_c}\frac{ \int^{\infty}_{-\infty}  
g_i(b_i)b_i^2db_i}{ \int^{\infty}_{-\infty}  
g_i(b_i)db_i} \nonumber \\
&+\frac{1}{4 \sigma^8_b} \sum_{i=1}^{n_c}  \left(\frac{\int^{\infty}_{-\infty}  
g_i(b_i)b_i^4db_i}{\int^{\infty}_{-\infty}  
g_i(b_i)db_i}-\frac{\left[\int^{\infty}_{-\infty}  
g_i(b_i)b_i^2db_i\right]^2}{\left[\int^{\infty}_{-\infty}  
g_i(b_i)db_i\right]^2} \right) \nonumber
\end{align}

{\setlength{\parskip}{10pt} \par \noindent  \textbf{Set 1 Exclusively}: Unique submatrices of $B_{\text{SSW}}$. 

For $B_{\text{SSW},31}$, we have 
\begin{align}
\sum_{i=1}^{n_c}\frac{\partial m_{{\text{SSW},i,p+2}}(\theta)}{\partial \beta^T}=&\sum_{i=1}^{n_c} \sum_{j=1}^{n_i} \frac{\partial}{\partial \beta^T}  \left\{\frac{Y_{ij}A_iS_{ij} \exp(D_{ij}(0)^T \beta + b_i)}{1+\exp(D_{ij}(0)^T \beta + b_i)} \right \} \\
&-\mu(1) \sum_{i=1}^{n_c} \sum_{j=1}^{n_i} \frac{\partial}{\partial \beta^T}  \left\{\frac{A_iS_{ij} \exp(D_{ij}(0)^T \beta + b_i)}{1+\exp(D_{ij}(0)^T \beta + b_i)} \right\} \nonumber \\
=&\sum_{i=1}^{n_c} \sum_{j=1}^{n_i}D_{ij}(0)^T\frac{Y_{ij}A_iS_{ij} \exp(D_{ij}(0)^T \beta + b_i)}{\left[1+\exp(D_{ij}(0)^T \beta + b_i)\right]^2} \nonumber \\
&-\mu(1) \sum_{i=1}^{n_c} \sum_{j=1}^{n_i}D_{ij}(0)^T \frac{A_iS_{ij} \exp(D_{ij}(0)^T \beta + b_i)}{\left[1+\exp(D_{ij}(0)^T \beta + b_i)\right]^2} \nonumber 
\end{align}
Analogously for $B_{\text{SSW},41}$, we have 
\begin{align}
\sum_{i=1}^{n_c}\frac{\partial m_{{\text{SSW},i,p+3}}(\theta)}{\partial \beta^T}=&\sum_{i=1}^{n_c} \sum_{j=1}^{n_i}D_{ij}(1)^T\frac{Y_{ij}(1-A_i)S_{ij} \exp(D_{ij}(1)^T \beta + b_i)}{\left[1+\exp(D_{ij}(1)^T \beta + b_i)\right]^2}  \\
&-\mu(0) \sum_{i=1}^{n_c} \sum_{j=1}^{n_i}D_{ij}(1)^T \frac{(1-A_i)S_{ij} \exp(D_{ij}(1)^T \beta + b_i)}{\left[1+\exp(D_{ij}(1)^T \beta + b_i)\right]^2} \nonumber
\end{align}
Finally for $B_{\text{SSW},33}$, we simply have 
\begin{equation}\sum_{i=1}^{n_c}\frac{\partial m_{{\text{SSW},i,p+2}}(\theta)}{\partial \mu(1)}=\sum_{i=1}^{n_c} \sum_{j=1}^{n_i} \frac{A_iS_{ij} \exp(D_{ij}(0)^T \beta + b_i)}{1+\exp(D_{ij}(0)^T \beta + b_i)}\end{equation}
and for $B_{\text{SSW},44}$,
\begin{equation}\sum_{i=1}^{n_c}\frac{\partial m_{{\text{SSW},i,p+3}}(\theta)}{\partial \mu(0)}=\sum_{i=1}^{n_c} \sum_{j=1}^{n_i} \frac{(1-A_i)S_{ij} \exp(D_{ij}(1)^T \beta + b_i)}{1+\exp(D_{ij}(1)^T \beta + b_i)}\end{equation}

{\setlength{\parskip}{10pt} \par \noindent  \textbf{Set 2 Exclusively}: Unique submatrices of $B_{\text{PSW}}$. 

Indeed, $B_{\text{PSW},31}=\sum_{i=1}^{n_c}\frac{\partial m_{{\text{PSW},i,p+2}}(\theta)}{\partial \beta^T}$ is the most challenging so let's start there. 
\begin{align}
B_{\text{PSW},31}=&\sum_{i=1}^{n_c} \sum_{j=1}^{n_i} \frac{\partial}{\partial \beta^T}  \left\{\frac{Y_{ij}A_iS_{ij} \exp(D_{ij}(0)^T \beta + b_i)}{1+\exp(D_{ij}(0)^T \beta + b_i)}\left[\frac{\exp(D_{ij}(1)^T \beta + b_i)}{1+\exp(D_{ij}(1)^T \beta + b_i)}\right]^{-1}\right \} \\
&-\sum_{i=1}^{n_c} \sum_{j=1}^{n_i} \frac{\partial}{\partial \beta^T}  \left\{ \frac{A_iS_{ij}\exp(D_{ij}(0)^T \beta + b_i)}{1+\exp(D_{ij}(0)^T \beta + b_i)}\left[\frac{\exp(D_{ij}(1)^T \beta + b_i)}{1+\exp(D_{ij}(1)^T \beta + b_i)}\right]^{-1}\right \}\nonumber \\
=&\sum_{i=1}^{n_c} \sum_{j=1}^{n_i} \left\{D_{ij}(0)^T\frac{Y_{ij}A_iS_{ij} \exp(D_{ij}(0)^T \beta + b_i)}{[1+\exp(D_{ij}(0)^T \beta + b_i)]^2}\left[\frac{\exp(D_{ij}(1)^T \beta + b_i)}{1+\exp(D_{ij}(1)^T \beta + b_i)}\right]^{-1}\right. \nonumber \\
&-\left.D_{ij}(1)^T\frac{Y_{ij}A_iS_{ij} \exp(D_{ij}(0)^T \beta + b_i)}{[1+\exp(D_{ij}(0)^T \beta + b_i)][\exp(D_{ij}(1)^T \beta + b_i)]} \right\} \nonumber \\
&-\mu(1)\sum_{i=1}^{n_c} \sum_{j=1}^{n_i} \left\{D_{ij}(0)^T\frac{A_iS_{ij} \exp(D_{ij}(0)^T \beta + b_i)}{[1+\exp(D_{ij}(0)^T \beta + b_i)]^2}\left[\frac{\exp(D_{ij}(1)^T \beta + b_i)}{1+\exp(D_{ij}(1)^T \beta + b_i)}\right]^{-1} \right. \nonumber \\
&-\left.D_{ij}(1)^T\frac{A_iS_{ij} \exp(D_{ij}(0)^T \beta + b_i)}{[1+\exp(D_{ij}(0)^T \beta + b_i)][\exp(D_{ij}(1)^T \beta + b_i)]} \right\} \nonumber
\end{align}
Lastly, we need just obtain $B_{\text{PSW},33}$, which is,
\begin{equation}\sum_{i=1}^{n_c}\frac{\partial  m_{{\text{PSW},i,p+2}}(\theta)}{\partial \mu(1)}=\sum_{i=1}^{n_c} \frac{A_iS_{ij}\exp(D_{ij}(0)^T \beta + b_i)}{1+\exp(D_{ij}(0)^T \beta + b_i)}\left[\frac{\exp(D_{ij}(1)^T \beta + b_i)}{1+\exp(D_{ij}(1)^T \beta + b_i)}\right]^{-1} \end{equation}
and for $B_{\text{PSW},44}$, we simply have 
\begin{equation}\sum_{i=1}^{n_c}\frac{\partial  m_{{\text{PSW},i,p+3}}(\theta)}{\partial \mu(0)}=\sum_{i=1}^{n_c} \sum_{j=1}^{n_i} (1-A_i)S_{ij}\end{equation}

\subsection{Outline of Numerical Integration Technique}\label{sec:app15}

We take account of all the 7 integrals forms listed below requiring approximation. As the integrals below are all (proportional to) Gaussian expectations they can be computed using the method of Gauss-Hermite quadrature\citep{liu1994note} as guided by previous literature.\citep{wu2019model,schafer2008average} The standard Gauss-Hermite quadrature approximation samples points that are roots of the Hermite polynomials, which are symmetric about 0. These may not cover higher density regions of the integrand. As such, Liu and Pierce\citep{liu1994note} outline a more principled way of selecting those points when computing intervals of the form $\int^{\infty}_{-\infty} h(t)dt$, where the ratio of $h(t)$ to the Gaussian kernel is moderately smooth. Indeed, the integrands below are all functions multiplied by $\exp\left(-\frac{b_i^2}{2\sigma^2_b}\right)$ that are evidently smooth. Additionally, it should be such that $h(t)$ is unimodal because it is approximated by 
\begin{equation}\int^{\infty}_{-\infty} h(t)dt \approx \sqrt{2}\widehat \sigma \sum_{i=1}^n w_i \exp\left(x_i^2\right)h(\widehat \mu + \sqrt{2}\widehat \sigma x_i)\end{equation}
where $\widehat \mu$ is the maximum of $h(t)$, $\widehat \sigma = \left(-\frac{\partial^2}{\partial t^2}\log h(t) \vert_{t=\widehat \mu}\right)^{-1/2}$, and $x_i$ are the roots of the $nth$ Hermite polynomial, $H_n(x)$, with weights $w_i=(\sqrt{\pi}2^{n-1}(n-1)!)(n\left\{H_{n-1}(x_i)\right\}^2)^{-1}$. We employed numerical methods available in standard statistical packages (such as \texttt{optimize()} in \texttt{R}) to find a maximum of $h(t)$. While this is not a guarantee of a global maximum (however, it can immediately be verified that 1-3 are indeed unimodal) if over a reasonable range of values the numerical algorithm failed to converge to a single maximum, the output (of our \texttt{R} package) reports an error; we did not explicitly encounter this issue for our simulations or data application. In such settings, we recommend using alternative methods for numerical integration. Let $\widehat g_i(b_i)$ denote replacing variables with observed values and parameters with their MLEs from Equation~\eqref{eq:kernel} as found in standard statistical software 
\begin{equation} \label{eq:kernelhat} \widehat g_i(b_i)=\exp \left( \sum_{j=1}^{n_i} \left[s_{ij}b_i-\log \left(1+\exp(d_{ij}^T \widehat \beta + b_i) \right)\right]-\frac{b_i^2}{2\widehat \sigma^2_b}\right)\end{equation}
For all $i$, we must compute integrals for the form:
\begin{enumerate} 
    \item $\int^{\infty}_{-\infty} \widehat g_i(b_i)db_i$
    \item $\int^{\infty}_{-\infty} \widehat g_i(b_i) b_i^2 db_i$
    \item $\int^{\infty}_{-\infty} \widehat g_i(b_i) b_i^4 db_i$
    \item $\int^{\infty}_{-\infty} \widehat g_i(b_i) \sum_{j=1}^{n_i}\frac{d_{ij}\exp(d_{ij}^T\widehat\beta+b_i)}{1+\exp(d_{ij}^T\widehat\beta+b_i)}db_i$
    \item $\int^{\infty}_{-\infty} \widehat g_i(b_i)b_i^2 \sum_{j=1}^{n_i}\frac{d_{ij} \exp(d_{ij}^T\widehat\beta+b_i)}{1+\exp(d_{ij}^T\widehat\beta+b_i)}db_i$
    \item $\int^{\infty}_{-\infty} \widehat g_i(b_i) \sum_{j=1}^{n_i}\frac{d_{ij}d_{ij}^T\exp(d_{ij}^T\widehat\beta+b_i)}{[1+\exp(d_{ij}^T\widehat\beta+b_i)]^2} db_i$
    \item $\int^{\infty}_{-\infty} \widehat g_i(b_i) \sum_{j=1}^{n_i}\frac{d_{ij}\exp(d_{ij}^T\widehat\beta+b_i)}{1+\exp(d_{ij}^T\widehat\beta+b_i)} \sum_{j'=1}^{n_i}\frac{d_{ij'}^T\exp(d_{ij'}^T\widehat\beta+b_i)}{1+\exp(d_{ij}^T\widehat\beta+b_i)}db_i$
\end{enumerate}

\subsection{Marginal Model Estimating Functions and Framework}\label{sec:app16}

In the case that we have a GLM with a cluster-robust variance expression, which we'll see is a GEE with independence working correlation (despite differing a priori assumptions), we have that \begin{equation}L(S|D,\beta)=\prod_{i=1}^{n_c} L(S_i|D_i, \beta)= \prod_{i=1}^{n_c} \left[\prod_{j=1}^{n_i} \frac{[\exp(D_{ij}^T \beta)]^{S_{ij}}}{1+\exp(D_{ij}^T \beta)}\right]\end{equation}
such that the log-likelihood is 
\begin{equation}l(S|D,\beta)=\sum_{i=1}^{n_c}\left ( \sum_{j=1}^{n_i} S_{ij}D_{ij}^T\beta-\sum_{j=1}^{n_i}\log[1+\exp(D_{ij}^T \beta)]\right)\end{equation}
where 
\begin{align}\frac{\partial l(S|D,\beta)}{\partial \beta}&=\sum_{i=1}^{n_c}\left ( \sum_{j=1}^{n_i} S_{ij}D_{ij}-\sum_{j=1}^{n_i}D_{ij}\frac{\exp(D_{ij}^T \beta)}{\log[1+\exp(D_{ij}^T \beta)]}\right)\\
&=\sum_{i=1}^{n_c}\left ( \sum_{j=1}^{n_i}D_{ij}\left\{ S_{ij}-\frac{\exp(D_{ij}^T \beta)}{1+\exp(D_{ij}^T \beta)}\right\}\right) \nonumber \\
&=\sum_{i=1}^{n_c}\left ( \sum_{j=1}^{n_i}\frac{D_{ij}\exp(D_{ij}^T \beta)}{[1+\exp(D_{ij}^T \beta)]^2}\frac{[1+\exp(D_{ij}^T \beta)]^2}{\exp(D_{ij}^T\beta)}\left\{ S_{ij}-\frac{\exp(D_{ij}^T \beta)}{1+\exp(D_{ij}^T \beta)}\right\}\right) \nonumber \\
&=\sum_{i=1}^{n_c} \left(\sum_{j=1}^{n_i} \frac{\partial \mu_{ij}(\beta)}{\partial \beta} v_{ij}(\beta)^{-1} \left\{ S_{ij}-\frac{\exp(D_{ij}^T \beta)}{1+\exp(D_{ij}^T \beta)}\right\}\right) \nonumber \\
&=\sum_{i=1}^{n_c} \mathbf{D}^T_i\mathbf{V}_i^{-1}(S_i-\mu_i(\beta)) \nonumber
\end{align}
where $\mu_i(\beta)=(\text{expit}(D_{i1}^T\beta),...,\text{expit}(D_{in_i}^T\beta))^T$, $\mathbf{D}_i=\frac{\partial \mu_i(\beta)}{\partial \beta^T}$, $\mathbf{V}_i=\text{Diag}(v_{i1}(\beta),...,v_{in_i}(\beta))$, and $S_i=(S_{i1},...,S_{in_i})^T$, which is precisely the same form as the GEE with independence working correlation. For the stacked estimating equations for Set 1, $\sum_{i=1}^{n_c}m_{\text{SSW},i}(\theta)$, we have:
\begin{equation} \label{eq:ms1imarg} m_{\text{SSW},i}(\theta)=\begin{pmatrix} \displaystyle \sum_{j=1}^{n_i}D_{ij}\left\{ S_{ij}-\frac{\exp(D_{ij}^T \beta)}{1+\exp(D_{ij}^T \beta)}\right\}\\  \sum_{j=1}^{n_i}Y_{ij}A_iS_{ij}p^0_{ij}(X_i,C_i)-\mu(1)\sum_{j=1}^{n_i}A_iS_{ij}p^0_{ij}(X_i,C_i)\\
 \sum_{j=1}^{n_i}Y_{ij}(1-A_i)S_{ij}p^1_{ij}(X_i,C_i)-\mu(0)\sum_{j=1}^{n_i}(1-A_i)S_{ij}p^1_{ij}(X_i,C_i)\end{pmatrix}\end{equation}
For stacked estimating equations for Set 2, $\sum_{i=1}^{n_c}m_{\text{PSW},i}(\theta)$ we have:

\begin{equation}\label{eq:ms2imarg} m_{\text{PSW},i}(\theta)=\begin{pmatrix} \displaystyle \sum_{j=1}^{n_i}D_{ij}\left\{ S_{ij}-\frac{\exp(D_{ij}^T \beta)}{1+\exp(D_{ij}^T \beta)}\right\} \\ 
\sum_{j=1}^{n_i} \frac{Y_{ij}A_iS_{ij} p^{0}_{ij}(X_i,C_i)}{ p^{1}_{ij}(X_i,C_i)}-\mu(1)\sum_{j=1}^{n_i}   \frac{A_iS_{ij}p^{0}_{ij}(X_i,C_i)}{p^{1}_{ij}(X_i,C_i)}\\
 \sum_{j=1}^{n_i}Y_{ij}(1-A_i)S_{ij}-\mu(0)\sum_{j=1}^{n_i}(1-A_i)S_{ij}\end{pmatrix} \end{equation}

 The asymptotic arguments and variance expressions proceed exactly as in the Supplemental Materials~\ref{sec:app13} and \ref{sec:app14}; these are straightforward and standard so we do not include their full work up. The estimators derived from the solutions to these estimating equations are $\widetilde \tau_{\text{SSW}}$ and $\widetilde \tau_{\text{PSW}}$. The top $1 \times 1$  block matrices for the asymptotic variances with these estimating functions $m_{(\cdot),i}(\theta)$ are where the GEE sandwich variance \citep{liang1986longitudinal} is recovered. 

\subsection{Data Generating Mechanism for Deterministic Monotonicity via Ordinal Representation}\label{sec:app17}

Suppose that we impose an ordinal structure on the 3 principal strata, removing the possibility of harmed patients. That is, let $G^{o}_{ij}=1$ if $G_{ij}=(0,0)$, $G^{o}_{ij}=2$ if $G_{ij}=(1,0)$, $G^{o}_{ij}=3$ if $G_{ij}=(1,1)$ such that $P(G^{o}_{ij} \in \{1,2,3\})=1$ for all $i,j$. Suppose that there is a latent continuous variable $G^{*}_{ij}$ and $\delta>0$ such that 
\begin{equation}G^{o}_{ij}= \begin{cases}
  1 & \text{if } G^{*}_{ij} \leq -\delta \\
  2 & \text{if }  -\delta < G^{*}_{ij} \leq 0 \\
  3 & \text{if } G^{*}_{ij} > 0
\end{cases}
\end{equation}
where $G^{*}_{ij}=D_{ij}^T \beta + b_i + \epsilon_{ij}$, $D_{ij}$ are the fixed effects excluding treatment, $b_i \sim N(0,\sigma^2_b)$, which are independent across $i$, and $\epsilon_{ij} \sim \text{Logistic}(0,1)$, which are independent across $i,j$. We know that if $G^{o}_{ij}=2$ or $G^{o}_{ij}=3$, then $S_{ij}(1)=1$ and similarly, if $G^{o}_{ij}=3$, then $S_{ij}(0)=1$. Therefore, we have that $S_{ij}(1)=1$ whenever $G^{*}_{ij} > -\delta$ and is 0 otherwise and $S_{ij}(0)=1$ whenever $G^{*}_{ij} > 0$ and is 0 otherwise. Moreover,
\begin{align}
\begin{split}
P(S_{ij}(1)=1|D_{ij},b_i) & =P(G^{*}_{ij} > -\delta|D_{ij},b_i)\\
& = P(D_{ij}^T \beta + b_i + \epsilon_{ij} > -\delta|D_{ij},b_i) \\
&= P(\epsilon_{ij} > -(\delta +D_{ij}^T \beta + b_i)) \\
&= P(\epsilon_{ij} \le \delta +D_{ij}^T \beta + b_i) \hspace{.2cm} \text{(by symmetry)}\\
&= \frac{1}{1+\exp\{-(\delta +D_{ij}^T \beta + b_i)\}}
\end{split}
\end{align}
Similarly,
$P(S_{ij}(0)=1|D_{ij},b_i)=\frac{1}{1+\exp\{-(D_{ij}^T \beta + b_i)\}}$. Thus, generating membership to these ordered principal strata according to these cut-points, $-\delta$ and 0, also implicitly generates survival under each treatment marginally with the positive effect of treatment 1 represented by $\delta$. Moreover, after random assignment, it would be correct model specification to use a GLMM (or if no clustering a GLM).

Note that for $j \ne j'$, \begin{equation}\text{Var}(G^{*}_{ij}|D_{ij})=\text{Var}(b_i)+\text{Var}(\epsilon_{ij})=\sigma^2_b+\pi^2/3=\text{Var}(G^{*}_{ij'}|D_{ij'})\end{equation} and \begin{equation}\text{Cov}(G^{*}_{ij},G^{*}_{ij'}|D_{ij},D_{ij'})=\text{Var}(b_i)=\sigma^2_b\end{equation} such that \begin{equation}
\text{Corr}(G^{*}_{ij},G^{*}_{ij'}|D_{ij},D_{ij'
})=\frac{\sigma^2_b}{\sigma^2_b+\pi^2/3}\end{equation} If $G^{o}_{ij}$ were binary, this is precisely how the ICC for binary outcomes is defined using standard statistical software as described earlier in Section 3 of the paper. 

Therefore, for this data generating process, we take the parameter values identical to those simulated in Section 3, but we replace the treatment effect with the following values: $\delta=\log(1.1), \log(1.25), \log(1.5)$. The percentage of always-survivors across these scenarios ranges from approximately $71\%$ when $\lambda=0.1$ and $68\%$ when $\lambda=0.3$. In Table~S\ref{tab:appmonres} of simulated results in Supplemental Materials~\ref{sec:appendix2}, we see no noticeable differences in the patterns that we remarked upon in Section 3 of the main paper.

\section{Additional Tables}\label{sec:appendix2}

\subsection{Table S1: Simulation Results for Additional Survival ICC Values}

\subsection{Table S2: Simulation Results for Negative Treatment Effects}

\subsection{Table S3: Simulation Results for Deterministic Monotonocity}

\subsection{Table S4: Comparison of Simulation Results and Computation Time for Asymptotic and Cluster Bootstrap Methods}

\setcounter{table}{0}

\begin{table}
\centering
\captionsetup{labelfont=bf,justification=raggedright,singlelinecheck=false,labelformat=addS}
    \renewcommand{\arraystretch}{0.8} 
    \caption{Additional scenarios ($\lambda=0.05,0.15$) for comparison of average point estimates and variance estimates as well as performance measures of bias and coverage ($\%$) for nominally $95\%$ z-intervals of the SSW and PSW estimators. Changes to cluster size $n_c$, treatment effects $\delta$, survival ICC $\lambda$, and modeling choices, GLMM (RE=T) vs. GLM (RE=F), are considered. Variances are computed using a degrees of freedom correction and are compared to the empirical variance (EV) in parentheses. Percentage of always-survivors range from $51\%$ to $65\%$, where larger values of $\delta$ and smaller $\lambda$ induce higher rates of this subgroup. 1000 simulations are run according to the proposed data generating mechanisms. $^{\dagger}$ indicates the setting of empirical monotonicity, and $^{*}$ indicates value is scaled up by $100$.} 
    \begin{tabular}{ccc|cc|cc|cc|cc}
    \toprule
        \multicolumn{11}{c}{$\lambda=0.05$}\\
    \midrule
        \multicolumn{3}{c|}{Scenarios} & \multicolumn{2}{c|}{Estimates} & \multicolumn{2}{c|}{Model Variance$^{*}$ (EV$^{*}$)} & \multicolumn{2}{c|}{Bias$^{*}$} & \multicolumn{2}{c}{Coverage}\\
        $\delta$ & $n_c$ & RE & SSW & PSW & SSW & PSW & SSW & PSW & SSW & PSW \\
    \midrule
        \multirow{6}{*}{0.0} & \multirow{2}{*}{30} & T & 1.568 & 1.564 & 2.7 (2.2) & 2.6 (2.2) & 0.0 & -0.4 & 95.7 & 96.2 \\
        & & F & 1.568 & 1.564 & 2.5 (2.1) & 2.5 (2.1) & -0.1 & -0.4 & 95.5 & 95.7 \\
        & \multirow{2}{*}{60} & T & 1.564 & 1.562 & 1.2 (1.1) & 1.1 (1.1) & -0.4 & -0.6 & 93.9 & 94.0 \\
        & & F & 1.565 & 1.562 & 1.1 (1.1) & 1.1 (1.1) & -0.3 & -0.6 & 93.8 & 94.0 \\
        & \multirow{2}{*}{90} & T & 1.566 & 1.563 & 0.8 (0.7) & 0.7 (0.7) & -0.2 & -0.5 & 95.7 & 95.4 \\
        & & F & 1.566 & 1.563 & 0.7 (0.7) & 0.7 (0.7) & -0.3 & -0.5 & 95.1 & 95.0 \\
    \midrule
        \multirow{6}{*}{0.2} & \multirow{2}{*}{30} & T & 1.563 & 1.561 & 2.7 (2.2) & 2.6 (2.2) & -0.5 & -0.7 & 96.0 & 95.9 \\
        & & F & 1.562 & 1.559 & 2.5 (2.1) & 2.5 (2.1) & -0.6 & -0.9 & 95.6 & 95.4 \\
        & \multirow{2}{*}{60} & T & 1.561 & 1.559 & 1.2 (1.1) & 1.1 (1.1) & -0.7 & -0.8 & 94.0 & 93.9 \\
        & & F & 1.559 & 1.557 & 1.1 (1.1) & 1.1 (1.1) & -0.9 & -1.1 & 93.3 & 93.6 \\
        & \multirow{2}{*}{90} & T & 1.562 & 1.560 & 0.8 (0.7) & 0.7 (0.7) & -0.6 & -0.8 & 95.7 & 95.8 \\
        & & F & 1.560 & 1.557 & 0.7 (0.7) & 0.7 (0.7) & -0.8 & -1.0 & 95.4 & 94.9 \\
    \midrule
        \multirow{6}{*}{1.6$^{\dagger}$} & \multirow{2}{*}{30} & T & 1.544 & 1.545 & 2.7 (2.1) & 2.7 (2.1) & -2.2 & -2.1 & 96.3 & 96.6 \\
        & & F & 1.539 & 1.538 & 2.5 (2.1) & 2.4 (2.1) & -2.7 & -2.8 & 95.2 & 95.3 \\
        & \multirow{2}{*}{60} & T & 1.542 & 1.543 & 1.2 (1.1) & 1.2 (1.1) & -2.4 & -2.3 & 94.8 & 94.9 \\
        & & F & 1.536 & 1.536 & 1.1 (1.1) & 1.1 (1.1) & -3.0 & -3.1 & 93.5 & 93.4 \\ 
        & \multirow{2}{*}{90} & T & 1.542 & 1.544 & 0.7 (0.7) & 0.7 (0.7) & -2.4 & -2.2 & 95.0 & 95.0 \\
        & & F & 1.536 & 1.536 & 0.7 (0.7) & 0.7 (0.7) & -3.0 & -3.0 & 93.7 & 93.7 \\
    \toprule
        \multicolumn{11}{c}{$\lambda=0.15$}\\
    \midrule
        \multirow{6}{*}{0.0} & \multirow{2}{*}{30} & T & 1.566 & 1.563 & 2.6 (2.5) & 2.5 (2.3) & -0.2 & -0.5 & 94.2 & 95.3 \\
        & & F & 1.566 & 1.563 & 2.4 (2.1) & 2.4 (2.1) & -0.2 & -0.5 & 94.9 & 95.2 \\
        & \multirow{2}{*}{60} & T & 1.563 & 1.562 & 1.1 (1.2) & 1.1 (1.1) & -0.4 & -0.6 & 92.6 & 93.6 \\
        & & F & 1.565 & 1.562 & 1.1 (1.0) & 1.1 (1.0) & -0.3 & -0.6 & 94.5 & 94.1 \\
        & \multirow{2}{*}{90} & T & 1.566 & 1.563 & 0.7 (0.7) & 0.7 (0.7) & -0.2 & -0.5 & 94.3 & 94.9 \\
        & & F & 1.566 & 1.563 & 0.7 (0.6) & 0.7 (0.6) & -0.2 & -0.5 & 95.6 & 95.5 \\
    \midrule
        \multirow{6}{*}{0.2} & \multirow{2}{*}{30} & T & 1.563 & 1.561 & 2.6 (2.4) & 2.5 (2.3) & -0.4 & -0.6 & 94.4 & 95.5 \\
        & & F & 1.559 & 1.556 & 2.3 (2.1) & 2.3 (2.1) & -0.9 & -1.1 & 95.1 & 95.0 \\
        & \multirow{2}{*}{60} & T & 1.560 & 1.559 & 1.1 (1.2) & 1.1 (1.1) & -0.7 & -0.8 & 92.9 & 92.9 \\
        & & F & 1.556 & 1.553 & 1.0 (1.0) & 1.0 (1.0) & -1.2 & -1.4 & 93.5 & 93.4 \\
        & \multirow{2}{*}{90} & T & 1.563 & 1.561 & 0.7 (0.7) & 0.7 (0.7) & -0.4 & -0.6 & 94.6 & 94.6 \\
        & & F & 1.557 & 1.555 & 0.7 (0.6) & 0.7 (0.6) & -1.0 & -1.3 & 95.2 & 95.5 \\
    \midrule
        \multirow{6}{*}{1.6$^{\dagger}$} & \multirow{2}{*}{30} & T & 1.543 & 1.544 & 2.6 (2.2) & 2.6 (2.2) & -2.4 & -2.2 & 95.1 & 95.3 \\
        & & F & 1.521 & 1.520 & 2.3 (2.0) & 2.3 (2.0) & -4.5 & -4.6 & 94.3 & 94.5 \\
        & \multirow{2}{*}{60} & T & 1.541 & 1.543 & 1.1 (1.1) & 1.1 (1.1) & -2.5 & -2.3 & 94.3 & 94.1 \\
        & & F & 1.517 & 1.517 & 1.0 (1.0) & 1.0 (1.0) & -4.9 & -4.9 & 90.5 & 90.4 \\
        & \multirow{2}{*}{90} & T & 1.544 & 1.545 & 0.7 (0.7) & 0.7 (0.7) & -2.2 & -2.1 & 94.7 & 94.9 \\
        & & F & 1.519 & 1.518 & 0.7 (0.6) & 0.7 (0.6) & -4.7 & -4.8 & 91.8 & 91.7 \\ 
    \bottomrule
    \end{tabular}
    \label{tab:app2}
\end{table}

\begin{table}
\centering
\captionsetup{labelfont=bf,justification=raggedright,singlelinecheck=false,labelformat=addS}
    \renewcommand{\arraystretch}{0.8} 
    \caption{Negative treatment effect scenarios for comparison of average point estimates and variance estimates as well as performance measures of bias and coverage ($\%$) for nominally $95\%$ z-intervals of the SSW and PSW estimators. Changes to cluster size $n_c$, survival ICC $\lambda$, and modeling choices, GLMM (RE=T) vs. GLM (RE=F), are considered. Variances are computed using a degrees of freedom correction and are compared to the empirical variance(EV) in parentheses. Percentage of always-survivors is about $48\%$ under all scenarios. 1000 simulations are run according to the proposed data generating mechanisms. $^{*}$ indicates value is scaled up by $100$.} 
    \begin{tabular}{ccc|cc|cc|cc|cc}
    \toprule
        \multicolumn{11}{c}{$\lambda=0.05$}\\
    \midrule
        \multicolumn{3}{c|}{Scenarios} & \multicolumn{2}{c|}{Estimates} & \multicolumn{2}{c|}{Model Variance$^{*}$ (EV$^{*}$)} & \multicolumn{2}{c|}{Bias$^{*}$} & \multicolumn{2}{c}{Coverage}\\
        $\delta$ & $n_c$ & RE & SSW & PSW & SSW & PSW & SSW & PSW & SSW & PSW \\
    \midrule
        \multirow{6}{*}{-0.2} & \multirow{2}{*}{30} & T & 1.571 & 1.566 & 2.7 (2.3) & 2.7 (2.2) & 0.2 & -0.3 & 96.0 & 96.1 \\
        & & F & 1.573 & 1.569 & 2.5 (2.2) & 2.5 (2.1) & 0.4 & 0.0 & 95.7 & 95.9 \\
        & \multirow{2}{*}{60} & T & 1.568 & 1.565 & 1.2 (1.2) & 1.2 (1.1) & 0.0 & -0.4 & 94.7 & 94.2 \\
        & & F & 1.571 & 1.568 & 1.1 (1.1) & 1.1 (1.1) & 0.3 & -0.1 & 94.9 & 94.2 \\
        & \multirow{2}{*}{90} & T & 1.569 & 1.565 & 0.8 (0.7) & 0.7 (0.7) & 0.1 & -0.4 & 95.6 & 95.2 \\ 
        & & F & 1.572 & 1.568 & 0.7 (0.7) & 0.7 (0.7) & 0.3 & 0.0 & 95.3 & 95.3 \\
    \toprule
        \multicolumn{11}{c}{$\lambda=0.1$}\\
    \midrule
        \multirow{6}{*}{-0.2} & \multirow{2}{*}{30} & T & 1.569 & 1.565 & 2.7 (2.4) & 2.6 (2.3) & 0.1 & -0.4 & 94.9 & 95.8 \\
        & & F & 1.574 & 1.570 & 2.5 (2.2) & 2.5 (2.1) & 0.5 & 0.2 & 95.3 & 95.6 \\
        & \multirow{2}{*}{60} & T & 1.567 & 1.564 & 1.2 (1.2) & 1.1 (1.1) & -0.1 & -0.4 & 94.1 & 94.2 \\
        & & F & 1.574 & 1.570 & 1.1 (1.1) & 1.1 (1.1) & 0.6 & 0.2 & 94.5 & 94.2 \\
        & \multirow{2}{*}{90} & T & 1.569 & 1.564 & 0.7 (0.7) & 0.7 (0.7) & 0.0 & -0.4 & 95.2 & 95.2 \\ 
        & & F & 1.573 & 1.570 & 0.7 (0.6) & 0.7 (0.6) & 0.5 & 0.1 & 95.4 & 95.3 \\
    \toprule
        \multicolumn{11}{c}{$\lambda=0.15$}\\
    \midrule
        \multirow{6}{*}{-0.2} & \multirow{2}{*}{30} & T & 1.570 & 1.566 & 2.6 (2.6) & 2.6 (2.3) & 0.2 & -0.2 & 93.8 & 94.7 \\
        & & F & 1.576 & 1.572 & 2.4 (2.2) & 2.4 (2.2) & 0.8 & 0.4 & 94.8 & 95.2 \\
        & \multirow{2}{*}{60} & T & 1.566 & 1.563 & 1.1 (1.3) & 1.1 (1.1) & -0.2 & -0.5 & 92.9 & 93.9 \\
        & & F & 1.574 & 1.570 & 1.1 (1.0) & 1.1 (1.0) & 0.6 & 0.2 & 94.4 & 94.4 \\
        & \multirow{2}{*}{90} & T & 1.568 & 1.564 & 0.7 (0.7) & 0.7 (0.7) & 0.0 & -0.4 & 94.2 & 95.0 \\ 
        & & F & 1.575 & 1.572 & 0.7 (0.6) & 0.7 (0.6) & 0.7 & 0.3 & 95.1 & 95.2 \\
    \toprule
        \multicolumn{11}{c}{$\lambda=0.3$}\\
    \midrule
        \multirow{6}{*}{-0.2} & \multirow{2}{*}{30} & T & 1.569 & 1.566 & 2.5 (2.9) & 2.5 (2.5) & 0.1 & -0.1 & 91.8 & 93.9 \\
        & & F & 1.579 & 1.576 & 2.3 (2.2) & 2.3 (2.1) & 1.2 & 0.8 & 94.2 & 94.6 \\
        & \multirow{2}{*}{60} & T & 1.564 & 1.564 & 1.1 (1.4) & 1.1 (1.2) & -0.3 & -0.3 & 90.7 & 92.6 \\
        & & F & 1.577 & 1.574 & 1.0 (1.0) & 1.0 (1.0) & 1.0 & 0.7 & 93.8 & 93.6 \\
        & \multirow{2}{*}{90} & T & 1.567 & 1.564 & 0.7 (0.8) & 0.7 (0.7) & -0.1 & -0.4 & 92.4 & 93.9 \\ 
        & & F & 1.577 & 1.574 & 0.7 (0.6) & 0.7 (0.6) & 1.0 & 0.6 & 95.1 & 95.1 \\
    \bottomrule
    \end{tabular}
    \label{tab:app2neg}
\end{table}

\begin{table}[ht]
\centering
\captionsetup{labelfont=bf,justification=raggedright,singlelinecheck=false,labelformat=addS}
    \renewcommand{\arraystretch}{0.8} 
    \caption{Deterministic monotonicity scenarios for comparison of average point estimates and variance estimates as well as performance measures of bias and coverage($\%$) for nominally $95\%$ z-intervals of the SSW and PSW estimators. Changes to cluster size $n_c$, treatment effects $\delta$, survival ICC $\lambda$, and modeling choices, GLMM (RE=T) vs. GLM (RE=F), are considered. Variances are computed using a degrees of freedom correction and are compared to the empirical variance(EV) in parentheses. 1000 simulations are run according to the proposed data generating mechanisms under deterministic monotonicity. $^{*}$ indicates value is scaled up by $100$.}
    \begin{tabular}{ccc|cc|cc|cc|cc}
    \toprule
        \multicolumn{11}{c}{$\lambda=0.1$}\\
    \midrule
        \multicolumn{3}{c|}{Scenarios} & \multicolumn{2}{c|}{Estimates} & \multicolumn{2}{c|}{Model Variance$^{*}$ (EV$^{*}$)} & \multicolumn{2}{c|}{Bias$^{*}$} & \multicolumn{2}{c}{Coverage}\\
        $\delta$ & $n_c$ & RE & SSW & PSW & SSW & PSW & SSW & PSW & SSW & PSW \\
    \midrule
        \multirow{6}{*}{0.1} & \multirow{2}{*}{30} & T & 1.575 & 1.574 & 2.7 (2.3) & 2.6 (2.2) &  1.0 & 0.8 & 95.2 & 95.5 \\
        & & F & 1.574 & 1.571 & 2.4 (2.0) & 2.4 (2.0) & 0.8 & 0.6 & 95.2 & 95.6 \\
        & \multirow{2}{*}{60} & T & 1.564 & 1.562 & 1.2 (1.1) & 1.1 (1.1) & -0.2 & -0.4 & 93.9 & 94.3 \\
        & & F & 1.562 & 1.559 & 1.1 (1.0) & 1.1 (1.0) & -0.4 & -0.6 & 94.7 & 94.5 \\
        & \multirow{2}{*}{90} & T & 1.568 & 1.566 & 0.7 (0.8) & 0.7 (0.8) & 0.3 & 0.0 & 93.4 & 93.6 \\
        & & F & 1.567 & 1.564 & 0.7 (0.7) & 0.7 (0.7) & 0.1 & -0.2 & 93.8 & 93.8 \\
    \midrule
       \multirow{6}{*}{0.2} & \multirow{2}{*}{30} & T & 1.573 & 1.572 & 2.7 (2.3) & 2.6 (2.2) & 0.7 & 0.6 & 95.4 & 94.9 \\
        & & F & 1.570 & 1.567 & 2.4 (2.0) & 2.4 (2.0) & 0.4 & 0.1 & 95.0 & 95.2 \\
        & \multirow{2}{*}{60} & T & 1.562 & 1.560 & 1.2 (1.1) & 1.1 (1.1) & -0.4 & -0.6 & 94.0 & 94.1 \\
        & & F & 1.558 & 1.555 & 1.1 (1.0) & 1.1 (1.0) & -0.8 & -1.0 & 93.9 & 93.9 \\
        & \multirow{2}{*}{90} & T & 1.566 & 1.564 & 0.7 (0.8) & 0.7 (0.8) & 0.0 & -0.1 & 94.0 & 94.1 \\
        & & F & 1.562 & 1.560 & 0.7 (0.7) & 0.7 (0.7) & -0.4 & -0.6 & 94.1 & 93.5 \\
    \midrule
        \multirow{6}{*}{0.4} & \multirow{2}{*}{30} & T & 1.571 & 1.570 & 2.7 (2.2) & 2.6 (2.1) & 0.5 & 0.5 & 95.3 & 95.1 \\
        & & F & 1.564 & 1.562 & 2.4 (2.0) & 2.4 (2.0)  & -0.2 & -0.4 & 95.1 & 94.8 \\
        & \multirow{2}{*}{60} & T & 1.559 & 1.558 & 1.2 (1.1) & 1.1 (1.1)  & -0.7 & -0.8 & 94.1 & 93.8 \\
        & & F & 1.552 & 1.550 & 1.1 (1.0) & 1.1 (1.0) & -1.4 & -1.6 & 93.9 & 94.0 \\
        & \multirow{2}{*}{90} & T & 1.564 & 1.563 & 0.7 (0.8) & 0.7 (0.8)  & -0.2 & -0.3 & 94.0 & 94.0 \\
        & & F & 1.557 & 1.554 & 0.7 (0.7) & 0.7 (0.7) & -0.9 & -1.1 & 93.9 & 94.0 \\ 
    \toprule
        \multicolumn{11}{c}{$\lambda=0.3$}\\
    \midrule
        \multirow{6}{*}{0.1} & \multirow{2}{*}{30} & T & 1.574 & 1.572 & 2.5 (2.7) & 2.5 (2.4)  & 0.8 & 0.7 & 92.5 & 93.3 \\
        & & F & 1.570 & 1.567 & 2.2 (2.0) & 2.3 (2.0) & 0.5 & 0.2 & 94.8 & 95.0 \\
        & \multirow{2}{*}{60} & T & 1.564 & 1.562 & 1.1 (1.3) & 1.1 (1.2) & -0.2 & -0.4 & 91.7 & 92.6 \\
        & & F & 1.561 & 1.558 & 1.0 (1.0) & 1.0 (1.0) & -0.5 & -0.7 & 94.5 & 94.3 \\
        & \multirow{2}{*}{90} & T & 1.568 & 1.567 & 0.7 (0.9) & 0.7 (0.8) & 0.3 & 0.1 & 90.4 & 91.6 \\
        & & F & 1.565 & 1.563 & 0.6 (0.7) & 0.6 (0.7) & 0.0 & -0.3 & 93.1 & 93.0 \\
    \midrule
        \multirow{6}{*}{0.2} & \multirow{2}{*}{30} & T & 1.573 & 1.571 & 2.5 (2.6) & 2.5 (2.4) & 0.7 & 0.6 & 92.8 & 93.2 \\
        & & F & 1.565 & 1.562 & 2.2 (2.0) & 2.2 (2.0) & -0.1 & -0.4 & 94.3 & 94.5 \\
        & \multirow{2}{*}{60} & T & 1.563 & 1.561 & 1.1 (1.3) & 1.1 (1.2) & -0.3 & -0.5 & 91.8 & 92.6 \\
        & & F & 1.554 & 1.552 & 1.0 (1.0) & 1.0 (1.0) & -1.1 & -1.4 & 94.0 & 94.3 \\
        & \multirow{2}{*}{90} & T & 1.567 & 1.566 & 0.7 (0.9) & 0.7 (0.8)  & 0.1 & 0.0 & 90.6 & 91.5 \\
        & & F & 1.559 & 1.557 & 0.6 (0.7) & 0.6 (0.7) & -0.7 & -0.9 & 93.2 & 93.4 \\
    \midrule
        \multirow{6}{*}{0.4} & \multirow{2}{*}{30} & T & 1.571 & 1.570 & 2.5 (2.6) & 2.4 (2.4)  & 0.6 & 0.5 & 93.1 & 93.4 \\
        & & F & 1.557 & 1.555 & 2.2 (2.0) & 2.2 (2.0) & -0.9 & -1.1 & 94.2 & 94.2 \\
        & \multirow{2}{*}{60} & T & 1.561 & 1.559 & 1.1 (1.2) & 1.1 (1.2)  & -0.5 & -0.6 & 92.3 & 92.7 \\
        & & F & 1.546 & 1.544 & 1.0 (1.0) & 1.0 (1.0) & -2.0 & -2.2 & 93.3 & 93.4 \\
        & \multirow{2}{*}{90} & T & 1.565 & 1.564 & 0.7 (0.9) & 0.7 (0.8) & 0.0 & -0.1 & 90.8 & 91.7 \\
        & & F & 1.550 & 1.548 & 0.6 (0.7) & 0.6 (0.7) & -1.5 & -1.7 & 93.4 & 93.6 \\
    \bottomrule
    \end{tabular}
    \label{tab:appmonres}
\end{table}

\begin{table}[ht]
    \centering
\captionsetup{labelfont=bf,justification=raggedright,singlelinecheck=false,labelformat=addS}
    \renewcommand{\arraystretch}{0.8} 
    \caption{Comparison of performance and computation time between asymptotic and resampling methods for variance estimation with $n_c=60$ clusters. Average variance estimates of SSW and PSW and coverage ($\%$) for nominally $95\%$ confidence intervals using the asymptotic sandwich variance (Asym.) and non-parametric cluster bootstrap method (Boot) with 250 replicates are juxtaposed. Confidence intervals are constructed using z-intervals and percentile intervals respectively. Estimation time in seconds represents the average total time (the implemented device took) to complete estimation of SACE and its variance for both SSW and PSW estimators. 1000 simulations are run according to described data generating mechanisms, where modeling choices, GLMM (RE=T) and GLM (RE=F), are considered. $^{\dagger}$ indicates the setting of empirical monotonicity.}
    \begin{tabular}{ccc|cc|cc|c}
        \toprule
        \multicolumn{8}{c}{$\lambda=0.1$}\\
        \midrule
        \multicolumn{3}{c|}{Scenarios} & \multicolumn{2}{c|}{Model Variance} & \multicolumn{2}{c|}{$\%$ Coverage} & Estimation \\
        $\delta$ & RE & Method & SSW & PSW & SSW & PSW & Time (s) \\
         \midrule
        \multirow{4}{*}{0.0} & \multirow{2}{*}{T} & Asym. & 0.012 & 0.011 & 93.1 & 93.9 & 32.5 \\
        &  & Boot & 0.011 & 0.010 & 92.6 & 92.8 & 340.3 \\
        & \multirow{2}{*}{F} & Asym. & 0.011 & 0.011 & 94.3 &  94.6 & 0.3 \\
        &  & Boot & 0.010 & 0.010 & 92.4 & 92.8 & 18.5 \\
        \multirow{4}{*}{0.2} & \multirow{2}{*}{T} & Asym. & 0.011 & 0.011 & 93.7 & 94.1 & 32.1 \\
        &  & Boot & 0.011 & 0.010 & 92.1 & 91.4 & 340.3 \\
        & \multirow{2}{*}{F} & Asym. & 0.011 & 0.011 & 94.5 & 94.6 & 0.3 \\
        &  & Boot & 0.010 & 0.010 & 93.0 & 93.1 & 22.3 \\
        \multirow{4}{*}{1.6$^{\dagger}$} & \multirow{2}{*}{T} & Asym. & 0.012 & 0.011 & 94.1 & 94.6 & 32.9 \\
        & & Boot & 0.010 & 0.010 & 92.9 & 93.1 & 340.3 \\
        & \multirow{2}{*}{F} & Asym. & 0.011 & 0.011 & 92.5 & 92.6 & 0.3 \\
        &  & Boot & 0.010 & 0.010 & 91.6 & 91.6 & 21.1 \\
        \toprule
        \multicolumn{8}{c}{$\lambda=0.3$}\\
        \midrule
        \multirow{4}{*}{0.0} & \multirow{2}{*}{T} & Asym. & 0.011 & 0.011 & 91.2 & 92.5 & 31.9 \\
        &  & Boot & 0.012 & 0.011 & 91.2 & 92.0 & 340.3 \\
        & \multirow{2}{*}{F} & Asym. & 0.010 & 0.010 & 94.2 & 93.5 & 0.3 \\
        &  & Boot & 0.009 & 0.009 & 92.3 & 92.6 & 19.5 \\
        \multirow{4}{*}{0.2} & \multirow{2}{*}{T} & Asym. & 0.011 & 0.011 & 91.8 & 92.4 & 33.6 \\
        &  & Boot & 0.011 & 0.011 & 91.5 & 92.7 & 340.3 \\
        & \multirow{2}{*}{F} & Asym. & 0.010 & 0.010 & 94.1 & 94.1 & 0.3 \\
        &  & Boot & 0.009 & 0.009 & 92.4 & 92.4 & 21.0 \\
        \multirow{4}{*}{1.6$^{\dagger}$} & \multirow{2}{*}{T} & Asym. & 0.011 & 0.011 & 92.8 & 93.1 & 33.9 \\
        & & Boot & 0.010 & 0.010 & 91.4 & 92.2 & 340.3 \\
        & \multirow{2}{*}{F} & Asym. & 0.010 & 0.010 & 89.6 & 89.2 & 0.3 \\
        &  & Boot & 0.009 & 0.009 & 89.0 & 89.0 & 20.3 \\
        \bottomrule
    \end{tabular}
    \label{tab:apptime}
\end{table}

\clearpage

\end{document}